\documentclass[10pt,twocolumn,amssymb,amsmath,aps,pra]{revtex4-1}
\usepackage{graphicx}

\begin{document}

\title{Lagrangian dynamics of the coupled field-medium state of light}
\date{July 30, 2019}
\author{Mikko Partanen}
\author{Jukka Tulkki}
\affiliation{Engineered Nanosystems Group, School of Science, Aalto University, P.O. Box 12200, 00076 Aalto, Finland}

\begin{abstract}
In the recently introduced mass-polariton (MP) theory of light [Phys.~Rev.~A 95, 063850 (2017)], the optical force of light drives in a medium forward an atomic mass density wave. In this work, we present the Lagrangian formulation of the MP theory starting directly from the principle of least action and the well-known Lagrangian densities of the electromagnetic field and the medium within the special theory of relativity. The Lagrangian densities and the resulting Euler-Lagrange equations lead directly and without any further \emph{postulates} to the unique expression of the optical Abraham force that dynamically couples the electromagnetic field and the medium in the MP theory of light. The field-medium coupling is \emph{symmetric and bi-directional and it fulfills the law of action and counteraction}. The coupled dynamical equations also enable the exact description of the very small kinetic energy of the medium as a part of the total energy of the coupled state of light. Thus, the Lagrangian formulation of the present work is a complementary approach to Lorentz covariance properties of the MP theory discussed in our recent work [Phys.~Rev.~A 99, 033852 (2019)]. We show how the coupled dynamical equations of the field and the medium can be solved analytically for a Gaussian light pulse. It is astonishing how the simple analytic results for the dynamical equations, the optical force, and the stress-energy-momentum tensor of the MP theory follow \emph{ab initio} from the Lagrangian densities that have been well known for almost a century.
\end{abstract}

\maketitle

\onecolumngrid
\vspace{-0.2cm}
\twocolumngrid

\section{Introduction}

The momentum of light and its formulation in the theory of electrodynamics in continuous media have been under an extensive debate since the early 20th century \cite{Leonhardt2006a,Cho2010,Barnett2010b,Bliokh2017a,Bliokh2017b,Pfeifer2007}. Defining and measuring the momentum of light has culminated in postulating separate Abraham and Minkowski stress-energy-momentum (SEM) tensors to describe the relations of the energy and momentum of light in condensed media \cite{Pfeifer2007,Penfield1967,Brevik1979,Kemp2017}. The fundamental question behind the Abraham-Minkowski controversy is how the momentum of light is split between the electromagnetic field and the atoms and how the atomic effects related to this coupling could be experimentally measured.

Very recently, the mass-polariton (MP) theory of light has provided a unique resolution to the Abraham-Minkowski controversy \cite{Partanen2017c,Partanen2017e,Partanen2019a}. The MP theory shows that light propagating in a medium must be described as a coupled state of the field and the medium. The key result of the MP theory is the atomic mass density wave (MDW) driven forward by the optical force. It is also an unavoidable consequence of the Lorentz transformation of the special theory of relativity and it is ultimately linked to the constant center of energy velocity of an isolated system. The MDW resulting from the coupling of the field and the medium does not only carry momentum, but also angular momentum. This has far-reaching consequences in the analysis of the angular momentum of light \cite{Partanen2018a} and the light-driven dynamics of atoms in optical fibers \cite{Partanen2018b}. The MP theory accounts not only the optical force, but also elastic forces between the atoms, which are displaced from their equilibrium positions by the optical force. Both these forces can be treated on equal footing in the dynamical equation of the medium enabling optoelastic continuum dynamics (OCD) simulations of the motion of atoms \cite{Partanen2017c}.

As the shift of atoms with the light-driven MDW is a classical mechanical quantity, it must be experimentally verifiable. Therefore, the MP theory provides a complementary approach to discover the momentum and angular momentum of light in different media. Thus, it may revive interest in experimental studies of the Abraham-Minkowski controversy \cite{Astrath2014,Ashkin1973,Casner2001,Pozar2018,Choi2017,Schaberle2019,Jones1954,Jones1978,Walker1975,She2008,Zhang2015,Campbell2005,Sapiro2009} and its relation to angular momentum transfer of light \cite{Allen1992b,Bliokh2015b,Bliokh2012,Bliokh2013a,Smirnova2018} in the presence of dielectric media.

In this work, we present the derivation of the MP theory directly from the principle of least action in the case of nondispersive transparent media. Complementary to the proof of the Lorentz covariance of the MP theory of light presented in Ref.~\cite{Partanen2019a}, the present work provides a solid classical field-theoretical basis for the MP theory. We show that the well-known physically intuitive Lagrangian densities of the field and the medium lead to intrinsic coupling of the Maxwell's equations of the field and Newton's equation of motion of the medium under the influence of the optical force. From these dynamical equations, we also derive the field and the medium parts of the total SEM tensor of the coupled system, which are in full agreement with the first-order approximations of these tensors presented in our previous work \cite{Partanen2019a}. While our previous work was based on taking the electromagnetic energy and momentum densities as given and neglected the extremely small kinetic energy of atoms resulting from the optical force, in the present work, we \emph{exactly} describe how small part of the total energy of light is carried as the kinetic energy of atoms. We also show that the field's share of the total energy of light is correspondingly reduced. The optical force density emerges from the Euler-Lagrange equations without further \emph{postulates} and becomes equal to the conventional Abraham force. Thus, our work also presents an \emph{ab initio} derivation of the Abraham force.

The present work is organized as follows: Section \ref{sec:action} presents the action principle and the Lagrangian densities of the field and the medium parts of the coupled dynamical system. We pay particular attention on the assumptions made on the Lagrangian densities. In Sec.~\ref{sec:EulerLagrange}, we write the Euler-Lagrange equations both for the electromagnetic field and the medium. This is followed by the formulation of the SEM tensors of the field and the medium in the laboratory frame in the MP theory of light in Sec.~\ref{sec:SEMtensors}. The SEM tensors are then generalized for an arbitrary inertial frame in Sec.~\ref{sec:gframe}. In Sec.~\ref{sec:solution}, we present an exact simultaneous solution of Maxwell's equations of the field and Newton's equation of the medium for a Gaussian plane wave pulse in a nondispersive medium. Brief comparison with our previous works is presented in Sec.~\ref{sec:comparison}. Finally, conclusions are drawn in Sec.~\ref{sec:conclusions}. Throughout our work, we include ample mathematical details of the derivations and, in the appendix, we present selected concepts of the special theory of relativity to facilitate the reading of our paper for less theoretical readers.

\section{\label{sec:action}Action principle and the Lagrangian density of the coupled field-medium system}

Following our previous work \cite{Partanen2019a}, we assume a homogeneous medium where material interfaces are not present. For simplicity, we also assume that the material is nondispersive, optically linear, and that there are no free charges and currents. The medium is also assumed to be lossless so that we neglect any optical absorption, but also any strain energies that are left in the medium by the light wave. The strain energies and elastic waves are known to be important in the description of the relaxation dynamics of the atomic displacements in the medium caused by the optical force  \cite{Partanen2017c,Partanen2017e,Partanen2018b}, but they are vanishingly small in comparison with the field energy. In realistic materials, the elastic energy density is also small with respect to the kinetic energy of atoms \cite{Partanen2017c}, and its negligence in the present work is effectively the same as the assumption of a medium for which the elastic constants are approximated to be zero. It is obvious that the description of elastic energies could be added in the present analysis, but this is left as a topic for future work. Even with these assumptions, the theory covers a broad range of optical phenomena and related applications in transparent materials. In a slightly more complex form, our concepts can be extended to account for optical absorption and dispersion.

Our approximations are mainly the same as those in our previous work \cite{Partanen2019a} with one exception. In our previous work, we assumed that the dynamical variables of the medium do not appear in the field part of the SEM tensor. In contrast, in the present work we account for the exact dependence of the Lagrangian density of the field on the four-velocity of the medium. Therefore, the present formulation of the MP theory of light is more general as it accounts for the kinetic energy of the MDW atoms, which have a small but nonzero velocity resulting from the optical force. We know from the previous OCD simulations \cite{Partanen2017c} that the kinetic energy of the MDW atoms resulting from the optical force is extremely small, although not exactly zero. Accounting for the four-velocity dependence of the Lagrangian density of the field allows us to analyze the dynamics of the action and the counteraction between the electromagnetic field and the medium through the optical force. Discussion of the specific terms that we include in the Lagrangian densities is presented in Sec.~\ref{sec:actionprinciple} below.

\subsection{\label{sec:actionprinciple}Action principle and the Lagrangian densities}

The Lagrangian formulation of the dynamics of the particles and fields is based on the principle of least action. This principle is defined by the statement that, for each system of particles and fields, there exists an action integral $S$, which obtains a minimum value for the true dynamics of the system, i.e., the variation $\delta S$ is zero. In terms of the Lagrangian density $\mathcal{L}$ of the system, the space-time action integral is written as \cite{Landau1989,Carroll2004}
\begin{equation}
 S=\int\mathcal{L}\sqrt{-g}\,d^4x,
 \label{eq:action}
\end{equation}
where $g$ is the determinant of the metric tensor.

In the present work, we study mass density perturbations of the medium generated by the optical force density associated with electromagnetic waves. However, these mass density perturbations are extremely small so that we can safely neglect any gravitational effects when describing the propagation of light at non-cosmological distance scales. Therefore, throughout the present work, we use the Minkowski metric tensor $g^{\alpha\beta}=\mathrm{diag}(1,-1,-1,-1)$, for which $g=-1$. Related to this, we neglect the well-known gravitational part $\mathcal{L}_\mathrm{grav}=-\frac{1}{2\kappa}R$ and the cosmological constant part $\mathcal{L}_\mathrm{cosm}=\frac{1}{\kappa}\Lambda_\mathrm{c}$ of the Lagrangian density, where $R$ is the Ricci scalar, $\Lambda_\mathrm{c}$ is the cosmological constant, and $\kappa=8\pi G/c^2$ is the Einstein constant, in which $G$ is the gravitational constant and $c$ is the speed of light in vacuum. Since we assume that the medium is nondispersive and that there are no free charges and currents, we neglect the explicit interaction part $\mathcal{L}_\mathrm{int}=-A_\alpha J^\alpha$ of the Lagrangian density, where $A_\alpha$ is the electromagnetic four-potential and $J^\alpha$ is the free electric four-current density.

With the assumptions above, we can write the Lagrangian density $\mathcal{L}$ of the coupled system of the electromagnetic field and the medium in an arbitrary inertial frame as a sum
\begin{equation}
 \mathcal{L} =\mathcal{L}_\mathrm{field}+\mathcal{L}_\mathrm{mat}.
 \label{eq:LagrangianDensityTotal}
\end{equation}
The well-known Lagrangian density $\mathcal{L}_\mathrm{field}$ of the electromagnetic field and the Lagrangian density $\mathcal{L}_\mathrm{mat}$ of the medium are given by \cite{Griffiths1998,Vanderlinde2004,Landau1984,Dirac1996} 
\begin{equation}
 \mathcal{L}_\mathrm{field}(\partial_\nu A_\mu,U_\mu)=-\frac{1}{4}F_{\mu\nu}\mathcal{D}^{\mu\nu},
 \label{eq:LagrangianDensityField}
\end{equation}
\begin{equation}
 \mathcal{L}_\mathrm{mat}(U_\mu)=\rho_0c\sqrt{U_\mu U^\mu}.
 \label{eq:LagrangianDensityMedium}
\end{equation}
Here $F_{\mu\nu}$ is the electromagnetic field tensor, $\mathcal{D}^{\mu\nu}$ is the electromagnetic displacement tensor, $\rho_0$ is the \emph{unperturbed rest} mass density of the medium, and $U^\mu$ is the four-velocity of the medium. The Greek indices range from 0 to 3 corresponding to the four components $(ct,x,y,z)$ of the Minkowski space-time. Below, in this work, the Latin indices $i$ and $j$ range from 1 to 3 corresponding to the three spatial components. Throughout the present work, we assume summation over repeated indices.

Since there is \emph{no direct coupling} term in the total Lagrangian density in Eq.~\eqref{eq:LagrangianDensityTotal}, the coupling of the field and the medium must be \emph{indirectly included} in the Lagrangian densities in Eqs.~\eqref{eq:LagrangianDensityField} and \eqref{eq:LagrangianDensityMedium}. This indirect coupling takes place through the four-velocity dependence of these Lagrangian densities as will be described below.

\subsection{Relation to previous Lagrangian formulations}

The Lagrangian densities of the electromagnetic field and the medium in Eqs.~\eqref{eq:LagrangianDensityField} and \eqref{eq:LagrangianDensityMedium} are both well known. However, in previous works, it is commonly assumed that the electromagnetic waves are not driving the medium when free charges and currents are absent. Thus, the four-velocity of the medium has been assumed as constant in an arbitrary inertial frame and zero in the laboratory frame, where the medium is assumed to be at rest before the arrival of electromagnetic waves. Accordingly, there has not been coupling between the well-known Lagrangian densities of the field and the medium in earlier works. Instead, one has typically tried to describe the interaction of the field and the medium with more complex heuristic Lagrangian densities to account for the medium part of the Lagrangian density \cite{Mikura1976,Obukhov2008,Ramos2015}. When we consider the influence of the optical force on the medium, the four-velocity of the medium is not generally constant but dynamically coupled to the values of the electric and magnetic fields through the space- and time-dependent optical force. The goal of the present work is to present the Lagrangian formalism of the coupled dynamics of the field and the medium starting from the well-known Lagrangian densities of the subsystems in Eqs.~\eqref{eq:LagrangianDensityField} and \eqref{eq:LagrangianDensityMedium}. Note that we do not assume any \emph{apriori} form of the optical force.

The field and the medium parts of the Lagrangian density are indirectly coupled to each other by the polarization and magnetization fields and the related optical force, which perturbs the mass density of the medium \cite{Partanen2017c}. This coupling is indirect as the total Lagrangian density of the coupled system of the field and the medium in Eq.~\eqref{eq:LagrangianDensityTotal} does not contain a separate term to describe this coupling. Due to the coupling of the field and the medium through the four-velocity of the medium and the related optical force, the field and medium parts of the Lagrangian density in Eqs.~\eqref{eq:LagrangianDensityField}--\eqref{eq:LagrangianDensityMedium} cannot be considered as Lagrangian densities of two separate isolated systems. The kinetic energy of atoms depends on the work done on them by the optical force. Thus, it is associated with the reduction of the field energy from its value in the case that the medium atoms would stay at rest. The four-velocity dependence of the Lagrangian density of the field, which leads to this indirect coupling of the Lagrangian densities, is described in detail in Sec.~\ref{sec:constitutive} below.

To schematically account for the field-medium coupling, the authors of some previous works \cite{Gordon1923,Leonhardt2006b,Leonhardt2010} have defined the so called electromagnetic Gordon metric in the medium. However, this approach is well known to lead to \emph{artificial} gravitational fields that are not physically true in the sense of the general theory of relativity \cite{Gordon1923}. This has also been briefly discussed in our previous work \cite{Partanen2019a}. Extensive review of previous efforts to account for the field-medium coupling is beyond the present work.

\subsection{\label{sec:quantities}Four-velocity, four-potential, and the electromagnetic field and displacement tensors}

In this subsection we briefly review the fundamental quantities of the space-time needed in the present work. These relations can be found in common textbooks \cite{Landau1987}. The four-velocity $U^\mu$ of the medium is given by $U^\mu=\frac{d}{d\tau}X^\mu=(\gamma c,\gamma v_\mathrm{a}^x,\gamma v_\mathrm{a}^y,\gamma v_\mathrm{a}^z)$, where $X^\mu$ is the position four-vector of the medium element, $v_\mathrm{a}^x$, $v_\mathrm{a}^y$, and $v_\mathrm{a}^z$ are components of the three-velocity vector $\mathbf{v}_\mathrm{a}=v_\mathrm{a}^i\mathbf{e}_i$ with length $v_\mathrm{a}=|\mathbf{v}_\mathrm{a}|$ and unit vectors $\mathbf{e}_i$, and $\gamma=1/\sqrt{1-v_\mathrm{a}^2/c^2}$ is the Lorentz factor.
The position four-vector of the medium element can be given as a function of the proper time $\tau$ as $X^\mu=[ct(\tau),x(\tau),y(\tau),z(\tau)]$.

The four-potential can be given in terms of the scalar potential $\phi$ and the vector potential $\mathbf{A}$ as $A^\alpha=(\phi/c,\mathbf{A})$. In terms of the four-potential, the covariant form of the electromagnetic field tensor $F_{\alpha\beta}$ is given by \cite{Jackson1999,Landau1989}
\begin{equation}
 F_{\alpha\beta}=\partial_\alpha A_\beta-\partial_\beta A_\alpha.
 \label{eq:electromagnetictensorfourpotential}
\end{equation}
In the contravariant matrix form, the electromagnetic field tensor $F^{\alpha\beta}$ can be written in Cartesian coordinates in terms of the $x$, $y$, and $z$ components of the electric field $\mathbf{E}$ and the magnetic flux density $\mathbf{B}$ as \cite{Jackson1999,Landau1989}
\begin{equation}
F^{\alpha\beta}=\left[\begin{array}{cccc}
0 & -E_x/c & -E_y/c & -E_z/c\\
E_x/c & 0 & -B_z & B_y\\
E_y/c & B_z & 0 & -B_x\\
E_z/c & -B_y & B_x & 0
\end{array}\right].
\label{eq:electromagnetictensor}
\end{equation}
This tensor is manifestly antisymmetric as it satisfies $F^{\alpha\beta}=-F^{\beta\alpha}$. The contravariant form of the electromagnetic displacement tensor $\mathcal{D}^{\alpha\beta}$, which combines the electric flux density $\mathbf{D}$ and the magnetic field $\mathbf{H}$, is given by \cite{Vanderlinde2004}
\begin{equation}
\mathcal{D}^{\alpha\beta}=\left[\begin{array}{cccc}
0 & -D_xc & -D_yc & -D_zc\\
D_xc & 0 & -H_z & H_y\\
D_yc & H_z & 0 & -H_x\\
D_zc & -H_y & H_x & 0
\end{array}\right].
\label{eq:displacementtensor}
\end{equation}
Like the electromagnetic field tensor in Eq.~\eqref{eq:electromagnetictensor}, the electromagnetic displacement tensor in Eq.~\eqref{eq:displacementtensor} is also antisymmetric satisfying $\mathcal{D}^{\alpha\beta}=-\mathcal{D}^{\beta\alpha}$. The four-velocity and four-potential dependencies of the electromagnetic displacement tensor are described in Sec.~\ref{sec:constitutive} below.

In the following derivations, we also use the Lorentz transformation matrix $\Lambda_{\;\,\beta}^\alpha$ corresponding to the atomic velocity $\mathbf{v}_\mathrm{a}$. Denoting $\mathbf{n}=\mathbf{v}_\mathrm{a}/v_\mathrm{a}$, the Lorentz boost $\Lambda_{\;\,\beta}^\alpha$ can be written in the matrix form as
\begin{equation}
 \Lambda_{\;\,\beta}^\alpha=\left[\begin{array}{cc}
  \gamma & -\gamma\frac{v_\mathrm{a}}{c}\mathbf{n}^T\\
  -\gamma\frac{v_\mathrm{a}}{c}\mathbf{n} & \mathbf{I}+(\gamma-1)\mathbf{n}\otimes\mathbf{n}
 \end{array}\right].
 \label{eq:Lorentzboost}
\end{equation}
The expression of $\Lambda_{\;\,\beta}^\alpha$ in terms of the spatial four-velocity components is obtained with substitutions $v_\mathrm{a}^i=U^i/\sqrt{1+u^2/c^2}$, where $u=|U^i\mathbf{e}_i|$.

\subsection{\label{sec:constitutive}Electromagnetic constitutive relations}

In this work, we consider propagation of light in a nondispersive isotropic medium. In the absence of light, the medium is assumed to be at rest in the \emph{laboratory frame} (L frame) excluding the possible thermal motion of atoms. When the optical force is present, the medium is generally put into position- and time-dependent motion in the L frame. We define the \emph{proper frame} as a frame that is attached to the medium element. Thus, the proper frame is not an inertial frame. However, at every point of space there is for every instance of time a separate \emph{local inertial frame} (A frame) comoving with the velocity of atoms with respect to the L frame. Since the formula of the optical force is unknown for the moment, the four-velocity of atoms due to the optical force is also unknown. The four-velocity of the medium is only obtained at a later stage as a result of the simultaneous solution of the Euler-Lagrange equations of the field and the medium. Note that, in contrast to the proper frame, the A frame is not attached to the pertinent medium element, and therefore, is not subject to acceleration. This momentarily comoving inertial frame is in common use in the theory of relativity. In addition, we define the \emph{general inertial frame} (G frame) as an arbitrary inertial frame that is generally in motion with respect to the medium.

\subsubsection{A frame}

We start by using the conventional constitutive relations in the A frame. These relations are given by \cite{Jackson1999,Griffiths1998}
\begin{equation}
 \mathbf{D}^\mathrm{(A)}=\varepsilon^\mathrm{(A)}\mathbf{E}^\mathrm{(A)},\hspace{0.7cm}\mathbf{B}^\mathrm{(A)}=\mu^\mathrm{(A)}\mathbf{H}^\mathrm{(A)},
 \label{eq:constitutive}
\end{equation}
where $\varepsilon^\mathrm{(A)}$ and $\mu^\mathrm{(A)}$ are the proper permittivity and permeability of the medium (i.e., defined in the A frame, where the medium element is momentarily at rest). The experimental permittivities and permeabilities given in the literature are based on the L frame measurements. These experimentally measured values become exactly equal to the proper material parameter values only at the weak field limit. For transparent solids, e.g, single-crystal silicon, the atomic velocities in the MDW are so small for all field strengths below the irradiation damage threshold that the difference between the L frame permittivities and permeabilities and the proper permittivities and permeabilities is certainly beyond the accuracy of the experimental measurements. The difference between the L frame and proper values of the permittivity and permeability is, however, theoretically extremely important as will become evident later in this work.

Using the constitutive relations in the A frame in Eq.~\eqref{eq:constitutive}, we obtain the relation between the electromagnetic field and displacement tensors in the A frame as
\begin{equation}
 (\mathcal{D}^\mathrm{(A)})_{\alpha\beta}=\frac{1}{\mu^\mathrm{(A)}}(d^\mathrm{(A)})_{\alpha\mu}(F^\mathrm{(A)})^{\mu\nu}(d^\mathrm{(A)})_{\nu\beta}
\label{eq:displacementtensorAframe}
\end{equation}
where $(d^\mathrm{(A)})_{\alpha\beta}=\mathrm{diag}[(n^\mathrm{(A)})^2,-1,-1,-1]$ is a diagonal matrix, in which $n^\mathrm{(A)}=c\sqrt{\varepsilon^\mathrm{(A)}\mu^\mathrm{(A)}}$ is the proper refractive index of the medium.

\subsubsection{G frame}

It is not practical to develop the Lagrangian formulation in the local inertial frame A. Instead, we derive the dynamical equations in the G frame to make their Lorentz covariance transparent. Therefore, we next present the relation of the tensors $F^{\mu\nu}$ and $\mathcal{D}_{\alpha\beta}$ in the G frame. In the following, the non-labeled tensors and four-vectors are given in the G frame (except in Sec.~\ref{sec:SEMtensors}, where we, for convenience, use no label for these quantities in the L frame). Using the Lorentz boost $\Lambda_{\;\,\beta}^\alpha$ from the general inertial frame to the A frame, given in Eq.~\eqref{eq:Lorentzboost}, we obtain $\mathcal{D}^{\alpha\beta}=g^{\alpha\chi}\Lambda_{\;\,\chi}^\mu(\mathcal{D}^\mathrm{(A)})_{\mu\nu}\Lambda_{\;\,\lambda}^\nu g^{\lambda\beta}$ and $(F^\mathrm{(A)})^{\alpha\beta}=\Lambda_{\;\,\gamma}^\alpha g^{\gamma\mu}F_{\mu\nu}g^{\nu\lambda}\Lambda_{\;\,\lambda}^\beta$. Here the metric tensor $g^{\alpha\beta}$ is only used for additionally raising and lowering tensor indices. Combining the two tensor transformations above with Eq.~\eqref{eq:displacementtensorAframe}, we then obtain
\begin{equation}
 \mathcal{D}^{\alpha\beta}=\frac{1}{\mu^\mathrm{(A)}}h^{\alpha\mu}F_{\mu\nu}h^{\nu\beta}.
 \label{eq:displacementtensorfourpotential}
\end{equation}
Here $h^{\alpha\mu}=g^{\alpha\chi}\Lambda_{\;\,\chi}^{\lambda} (d^\mathrm{(A)})_{\lambda\gamma}\Lambda_{\;\,\delta}^\gamma g^{\delta\mu}$ is a symmetrix matrix that is formed from the Lorentz boosts, the metric tensor, and the diagonal matrix $(d^\mathrm{(A)})_{\alpha\beta}$. Using the definitions of the mentioned quantities above, one then finds an explicit expression of $h^{\alpha\beta}$ in terms of the components of the four-velocity of the medium as
\begin{equation}
 h^{\alpha\beta}=g^{\alpha\beta}+\frac{(n^\mathrm{(A)})^2-1}{c^2}U^\alpha U^\beta.
\label{eq:htensor}
\end{equation}

Also, note that the definitions of $F^{\alpha\beta}$ and $\mathcal{D}^{\alpha\beta}$ in Eqs.~\eqref{eq:electromagnetictensorfourpotential} and \eqref{eq:displacementtensorfourpotential} and the constitutive relations of the A frame in Eq.~\eqref{eq:constitutive} can be used to derive the generalized form of the constitutive relations for the G frame. These generalized constitutive relations, given by $\mathcal{D}^{\alpha\beta}U_\beta=c^2\varepsilon^\mathrm{(A)} F^{\alpha\beta}U_\beta$ and $\star\mathcal{D}^{\alpha\beta}U_\beta=\frac{1}{\mu^\mathrm{(A)}}\star F^{\alpha\beta}U_\beta$, where $\star$ denotes the Hodge dual \cite{Griffiths1998}, are well known and in agreement with our Eqs.~\eqref{eq:displacementtensorfourpotential} and \eqref{eq:htensor}.

\subsubsection{L frame}

Instead of using the constitutive relations in Eq.~\eqref{eq:constitutive} in the A frame,
it would be tempting to use these relations in the L frame as $\mathbf{D}^\mathrm{(L)}=\varepsilon^\mathrm{(L)}\mathbf{E}^\mathrm{(L)}$ and $\mathbf{B}^\mathrm{(L)}=\mu^\mathrm{(L)}\mathbf{H}^\mathrm{(L)}$, where $\varepsilon^\mathrm{(L)}$ and $\mu^\mathrm{(L)}$ are the permittivity and permeability in the L frame. Then, we would have $(\mathcal{D}^\mathrm{(L)})_{\alpha\beta}=\frac{1}{\mu^\mathrm{(L)}}(d^\mathrm{(L)})_{\alpha\mu}(F^\mathrm{(L)})^{\mu\nu}(d^\mathrm{(L)})_{\nu\beta}$, where $(d^\mathrm{(L)})_{\alpha\beta}=\mathrm{diag}[(n^\mathrm{(L)})^2,-1,-1,-1]$ is a diagonal matrix and $n^\mathrm{(L)}=c\sqrt{\varepsilon^\mathrm{(L)}\mu^\mathrm{(L)}}$ is the refractive index of the medium in the L frame. In these relations, the dependence on the four-velocity of the medium is only implicitly present through the material parameters of the L frame, which are related to the material parameters of the A frame as presented in Appendix \ref{apx:materialparameters}. The implicit dependence of the material parameters on the four-velocity of the medium is not convenient regarding the derivation of the dynamical equations of the medium. Thus, in the present work, we use the expression of $\mathcal{D}^{\alpha\beta}$ in Eq.~\eqref{eq:displacementtensorfourpotential}, where the four-velocity dependence is explicitly present through Eq.~\eqref{eq:htensor}. The two approaches are, however, equivalent.

\section{\label{sec:EulerLagrange}Euler-Lagrange equations}

\subsection{\label{sec:EulerLagrangeMedium}Euler-Lagrange equations for the medium}

Next, we derive the dynamical equations of the medium in the G frame. Keeping the four-potential and the proper permittivity and permeability constant and varying the action in Eq.~\eqref{eq:action} with respect to the position four-vector of the medium element gives the Euler-Lagrange equations as
\begin{equation}
 \frac{\partial\mathcal{L}}{\partial X_\alpha}-\frac{d}{d\tau}\Big(\frac{\partial\mathcal{L}}{\partial U_\alpha}\Big)=0.
\label{eq:eulerlagrangemat1}
\end{equation}
Using the Lagrangian densities of the electromagnetic field and the medium, given in Eqs.~\eqref{eq:LagrangianDensityField} and \eqref{eq:LagrangianDensityMedium}, we obtain the following relations
\begin{align}
 &\frac{\partial\mathcal{L}_\mathrm{field}}{\partial X_\alpha}=0,
 \hspace{0.4cm}\frac{\partial\mathcal{L}_\mathrm{mat}}{\partial X_\alpha}=0,
 \hspace{0.4cm}\frac{\partial\mathcal{L}_\mathrm{mat}}{\partial U_\alpha}=\rho_0U^\alpha,\nonumber\\[5pt]
 &\frac{\partial\mathcal{L}_\mathrm{field}}{\partial U_\alpha}=
 -\frac{1}{\gamma c}\frac{|(F_{\;\,\mu}^{i}\mathcal{D}_{\;\,0}^\mu-\mathcal{D}_{\;\,\mu}^{i} F_{\;\,0}^\mu)\mathbf{e}_i|}{|U^j\mathbf{e}_j|}U^\alpha\nonumber\\
 &\hspace{1.05cm}=\begin{cases}-\frac{1}{\gamma c}(F_{\;\,\mu}^{\alpha}\mathcal{D}_{\;\,0}^\mu-\mathcal{D}_{\;\,\mu}^{\alpha} F_{\;\,0}^\mu),\text{ for $\alpha=1,2,3,$}\\[5pt]
 -\frac{1}{\gamma c}\frac{|(F_{\;\,\mu}^{i}\mathcal{D}_{\;\,0}^\mu-\mathcal{D}_{\;\,\mu}^{i} F_{\;\,0}^\mu)\mathbf{e}_i|}{|U^j\mathbf{e}_j|}U^\alpha,\text{ for $\alpha=0$.}
 \end{cases}
 \label{eq:derivativerelations1}
\end{align}

Unlike most four-vectors, the four-velocity $U^\alpha$ has only three independent components instead of four. This follows from the fact that the time component, $\gamma c$, is a function of the space components. Therefore, the mentioned dependence between the four-velocity components must be accounted for when differentiating the Lagrangian densities with respect to $U^\alpha$ to obtain the results in Eq.~\eqref{eq:derivativerelations1}. However, note that when the four-velocity is multiplied with $\rho_0$, which is a Lorentz scalar, we get the Lagrangian momentum density four-vector $P^\mu=\rho_0 U^\mu$, which has four independent components. Thus, effectively the time component, $\gamma c$, combines with $\rho_0$ to make the fourth independent component.

In the Lagrangian density of the medium, $\rho_0$ and $U^\alpha$ are not independent as described above. To obtain the expression of $\partial\mathcal{L}_\mathrm{mat}/\partial U_\alpha$ in Eq.~\eqref{eq:derivativerelations1}, this fact can be circumvented by utilizing the Lagrangian momentum density four-vector $P^\mu=\rho_0 U^\mu$. Then the Lagrangian density of the medium can be written as $\mathcal{L}_\mathrm{mat}=c\sqrt{P_\mu P^\mu}$ \cite{Dirac1996}. Thus, $\partial\mathcal{L}_\mathrm{mat}/\partial U_\alpha=(\partial\mathcal{L}_\mathrm{mat}/\partial P_\beta)(\partial P_\beta/\partial U_\alpha)$, and using the relations $\partial\mathcal{L}_\mathrm{mat}/\partial P_\beta=cP^\beta/\sqrt{P_\mu P^\mu}=U^\beta$ and $\partial P_\beta/\partial U_\alpha=\rho_0\delta_\beta^\alpha$, we then obtain the result $\partial\mathcal{L}_\mathrm{mat}/\partial U_\alpha=\rho_0 U^\alpha$ as given in Eq.~\eqref{eq:derivativerelations1}.

The derivation of $\partial\mathcal{L}_\mathrm{field}/\partial U_\alpha$ in Eq.~\eqref{eq:derivativerelations1} is technical, but still straightforward using the relations given in the sections above. For the spatial four-velocity components, the expression of $\partial\mathcal{L}_\mathrm{field}/\partial U_i$ is obtained by using the definition of $\mathcal{D}^{\alpha\beta}$ in terms of $h^{\alpha\beta}$, given in Eq.~\eqref{eq:displacementtensorfourpotential}, and the expression of $h^{\alpha\beta}$, given in Eq.~\eqref{eq:htensor}. Due to the fact that the four-velocity has only three independent components as discussed above, in this calculation, the time component of the four-velocity must be presented as a function of the space components as $U_0=\sqrt{c^2+u^2}$. For the temporal four-velocity component, the expression for $\partial\mathcal{L}_\mathrm{field}/\partial U_0$ is obtained from the chain rule as $\partial\mathcal{L}_\mathrm{field}/\partial U_0=(\partial\mathcal{L}_\mathrm{field}/\partial u)(\partial u/\partial U_0)$.

Thus, using the derivatives of the Lagrangian densities in Eq.~\eqref{eq:derivativerelations1}, the Euler-Lagrange equations in Eq.~\eqref{eq:eulerlagrangemat1} can be written as
\begin{equation}
 \frac{d}{d\tau}(\rho_0U^\alpha)
=-\frac{d}{d\tau}\Big(\frac{\partial\mathcal{L}_\mathrm{field}}{\partial U_\alpha}\Big),
\label{eq:eulerlagrangemat}
\end{equation}
Since the left hand side of Eq.~\eqref{eq:eulerlagrangemat} is the time derivative of the momentum density of the medium and the right hand is the force density due to the fields, Eq.~\eqref{eq:eulerlagrangemat} is essentially Newton's equation of motion for the medium in the presence of field-induced forces.

However, note that the proper time along a timelike world line of the medium element is, by definition, the time that is measured by a clock following that world line. Therefore, the proper time used in Eq.~\eqref{eq:eulerlagrangemat} is not exactly the same as the time measured by an observer in the L frame, where the medium element is moving due to the optical force. The same naturally applies to the general G frame, where the medium element can additionally have velocity components independent of the optical force. Using the well-known relation between the differentials of the proper time and the coordinate time, given by $dt=\gamma d\tau$, we can write Newton's equation of motion in Eq.~\eqref{eq:eulerlagrangemat} in the coordinate form in the G frame as
\begin{equation}
 \gamma\frac{d}{dt}\Big(\frac{\rho_0^\mathrm{(G)} U^\alpha}{\gamma^2}\Big)
=-\gamma\frac{d}{dt}\Big(\frac{\partial\mathcal{L}_\mathrm{field}}{\partial U_\alpha}\Big),
\label{eq:eulerlagrangematL}
\end{equation}
Here $\rho_0^\mathrm{(G)}=\gamma^2\rho_0$ is the \emph{unperturbed} mass density of the medium in the G frame, where the unperturbed rest mass density $\rho_0=\rho_0^\mathrm{(A)}$ of the A frame is moving. Note that all \emph{unlabeled mass densities} throughout this work are defined in the A frame, which is the local rest frame of the medium. The two $\gamma$ factors in $\rho_0^\mathrm{(G)}$ originate from the Lorentz contraction and the kinetic energy of $\rho_0$. The changes in the instantaneous atomic mass density of the medium, which follow from the space- and time-dependent atomic displacements due to the optical force, are not included in $\rho_0^\mathrm{(G)}$. Thus, in the special case of the L frame, $\rho_0^\mathrm{(L)}$ is equal to the \emph{rest} mass density of the medium in the \emph{absence} of the electromagnetic field. Below, in Secs.~\ref{sec:SEMtensorMedium} and \ref{sec:solution}, we will present in detail how $\rho_0^\mathrm{(L)}$ is related to the true perturbed mass density of the medium, which includes the effects of the atomic displacements by the optical force. Note that the $\gamma$ factors are extremely close to unity in the L frame, where the velocity of atoms is very much smaller than the velocity of light. However, these factors are still important in the exact theoretical description.

It is essential to note that the optical force on the right hand side of Eqs.~\eqref{eq:eulerlagrangemat} and \eqref{eq:eulerlagrangematL} describes how the state of motion of a single medium element changes in the course of time. Another approach would be to study how the flow velocity of medium elements passing a fixed spatial position is changing as a function of time. These two approaches are well-known as the Lagrangian and Eulerian descriptions of the flow velocity field of the medium \cite{Landau1987}. Conservation laws are conveniently expressed in the Eulerian form. Thus, we transform Eq.~\eqref{eq:eulerlagrangematL} into the equivalent Eulerian form to derive the SEM tensor of the medium and the related conservation laws. This will be described in Sec.~\ref{sec:SEMtensorMedium}.

\subsection{\label{sec:EulerLagrangeField}Euler-Lagrange equations for the electromagnetic field}

In this section, we briefly review the well-known derivation of the Maxwell's equations from the Lagrangian density in the G frame. Keeping the four-velocity of the medium and the proper permittivity and permeability constant and varying the action in Eq.~\eqref{eq:action} with respect to the four-potential gives the Euler-Lagrange equations as \cite{Landau1989}
\begin{equation}
 \frac{\partial\mathcal{L}}{\partial A_\mu}-\partial_\lambda\Big[\frac{\partial\mathcal{L}}{\partial(\partial_\lambda A_\mu)}\Big]=0.
 \label{eq:EulerLagrange1}
\end{equation}
Using the expressions of the electromagnetic and displacement tensors in terms of the four-potential in Eqs.~\eqref{eq:electromagnetictensorfourpotential} and \eqref{eq:displacementtensorfourpotential}, we obtain the well-known relations
\begin{align}
 \frac{\partial\mathcal{L}_\mathrm{field}}{\partial A_\beta} &=0,
\hspace{0.4cm}\frac{\partial\mathcal{L}_\mathrm{field}}{\partial(\partial_\alpha A_\beta)}=-\mathcal{D}^{\alpha\beta},\nonumber\\
 \frac{\partial\mathcal{L}_\mathrm{mat}}{\partial A_\beta} &=0,
\hspace{0.4cm}\frac{\partial\mathcal{L}_\mathrm{mat}}{\partial(\partial_\alpha A_\beta)}=0.
\label{eq:EulerLagrangeParts}
\end{align}
Using the relations in Eq.~\eqref{eq:EulerLagrangeParts}, the Euler-Lagrange equations in Eq.~\eqref{eq:EulerLagrange1} can be written as \cite{Landau1989}
\begin{equation}
\partial_\alpha\mathcal{D}^{\alpha\beta}=0. %J_\mathrm{f}^\beta.
 \label{eq:EulerLagrangeD}
\end{equation}
This is known as the Gauss-Ampere law \cite{Vanderlinde2004} and it is written in our case by assuming the absence of free charges and currents. It is a combination of two of the four Maxwell's equations, namely Gauss's and Amp\`ere's laws. 

For completeness, we also note that the remaining two Maxwell's equations, Faraday's law of induction and Gauss's law for magnetism, can be written together as the Gauss-Faraday law, given by \cite{Landau1989}
\begin{equation}
 \partial_\alpha(\textstyle\frac{1}{2}\epsilon^{\alpha\beta\mu\nu}F_{\mu\nu})=0.
 \label{eq:Maxwell2}
\end{equation}
Here $\epsilon^{\alpha\beta\mu\nu}$ is the Levi-Civita symbol. Equation \eqref{eq:Maxwell2} follows directly from the definition of the electromagnetic field tensor in terms of the four-potential in Eq.~\eqref{eq:electromagnetictensorfourpotential} through the Bianchi identity in the same way as in vacuum \cite{Bliokh2013b,Landau1989}.

The form of Maxwell's equations derived above is standard, but these equations are coupled to the dynamical state of the medium via the atomic velocity and its relation to the permittivity and permeability of the medium. Thus, the Euler-Lagrange equations of the field in \eqref{eq:EulerLagrange1} and the Euler-Lagrange equation of the medium in Eq.~\eqref{eq:eulerlagrangemat} are bi-directionally coupled dynamical equations. In the general case, these coupled equations can be solved only numerically to obtain a self-consistent solution. However, there are some special cases for which the exact analytical solutions can be obtained, see Sec.~\ref{sec:solution}.

\section{\label{sec:SEMtensors}SEM tensors of the electromagnetic field and the medium in the L frame}

In previous sections, the physics of light is described by the physical variables of the field and the medium and their dynamical equations. Next, we formulate the SEM tensors of the electromagnetic field and the medium based on the Euler-Lagrange equations of Sec.~\ref{sec:EulerLagrange}. The conventional definition of the SEM tensor in the Minkowski space-time is given by \cite{Landau1989,Jackson1999,Misner1973}
\begin{equation}
 T^{\alpha\beta}=
 \left[\begin{array}{cc}
  W & c\mathbf{G}^T\\
  c\mathbf{G} & \boldsymbol{\mathcal{T}}\\
 \end{array}\right]
 =\left[\begin{array}{cccc}
  W & cG^x & cG^y & cG^z\\
  cG^x & \mathcal{T}^{xx} & \mathcal{T}^{xy} & \mathcal{T}^{xz}\\
  cG^y & \mathcal{T}^{yx} & \mathcal{T}^{yy} & \mathcal{T}^{yz}\\
  cG^z & \mathcal{T}^{zx} & \mathcal{T}^{zy} & \mathcal{T}^{zz}
 \end{array}\right],
 \label{eq:emt}
\end{equation}
where $W$ is the energy density, $\mathbf{G}=(G^x,G^y,G^z)$ is the momentum density, and $\boldsymbol{\mathcal{T}}$ is the stress tensor.  The stress tensor components $\mathcal{T}^{ij}$ describe the flux of $i$th component of linear momentum across the $x^j$ surface.

For an isolated system, the SEM tensor serves as a concise all-in-one presentation of the independent physical quantities of the system and the related conservation laws. In terms of the SEM tensor, the conservation laws of energy, momentum, and angular momentum in the Minkowski space-time are compactly written as $\partial_\beta T^{\alpha\beta}=0$ and $T^{\alpha\beta}=T^{\beta\alpha}$ \cite{Jackson1999}. It is also well known that the total SEM tensor of a closed system is in general related to the curvature of the space-time through Einstein's field equations.

From the conservation laws above, it follows that the SEM tensor of a closed system is automatically symmetric. The definition of the SEM tensor in Eq.~\eqref{eq:emt} can also be used for interacting subsystems. However, the conservation laws are not separately fulfilled for subsystems, and therefore, the SEM tensors of interacting subsystems do not need to be symmetric. Below, we derive the SEM tensors of the field and medium subsystems directly from the Euler-Lagrange equations using the definition of the SEM tensor in Eq.~\eqref{eq:emt}. However, we do not make any further assumptions on the symmetry of these SEM tensors below.

\subsection{\label{sec:SEMtensorMedium}SEM tensor of the medium}

In this subsection, we derive the SEM tensor of the medium in the L frame. The dynamical Euler-Lagrange equations for the medium were given in the Lagrangian form in Eqs.~\eqref{eq:eulerlagrangemat} and \eqref{eq:eulerlagrangematL} that are equivalent. As discussed in Sec.~\ref{sec:EulerLagrangeMedium}, these dynamical equations must be transformed into the Eulerian form to enable writing the SEM tensor and the related conservation laws.

The coordinate form of the Euler-Lagrange equations in the G frame, given in Eq.~\eqref{eq:eulerlagrangematL}, is the starting point of our derivation in this section. To transform Eq.~\eqref{eq:eulerlagrangematL} into the Eulerian form, we first define two key quantities of the MP theory of light. These quantities are the \emph{true perturbed rest} mass density of the medium denoted by $\rho_\mathrm{a}=\rho_\mathrm{a}^\mathrm{(A)}$ and the \emph{excess} mass density of the medium in the G frame denoted by $\rho_\mathrm{MDW}^\mathrm{(G)}$. The relation of these quantities to the unperturbed rest mass density $\rho_0=\rho_0^\mathrm{(A)}$ and the four-velocity $U^\alpha$ of the medium is unambiguously defined in the L frame by the two equations, given by
\begin{equation}
 \gamma^2\rho_\mathrm{a}\frac{dU^\alpha}{dt}=\gamma^2\rho_0\frac{\partial U^\alpha}{\partial t},
\label{eq:mprelation1}
\end{equation}
\begin{equation}
 \gamma^2\rho_\mathrm{a}=\gamma^2\rho_0+\rho_\mathrm{MDW}^\mathrm{(L)}.
\label{eq:mprelation2}
\end{equation}

Using Eq.~\eqref{eq:mprelation2}, we rewrite the left hand side of the dynamical equation of the medium in Eq.~\eqref{eq:eulerlagrangematL} as
\begin{align}
 &\gamma\frac{d}{dt}\Big(\frac{\rho_0^\mathrm{(L)}U^\alpha}{\gamma^2}\Big)\nonumber\\
 &=\gamma\frac{d(\rho_\mathrm{a}U^\alpha)}{dt}
-\gamma\frac{d}{dt}\Big(\frac{\rho_\mathrm{MDW}^\mathrm{(L)}U^\alpha}{\gamma^2}\Big)\nonumber\\
 &=\gamma^2\rho_\mathrm{a}\frac{d^2X^\alpha}{dt^2}
+U^\alpha\frac{d(\gamma\rho_\mathrm{a})}{dt}
-\gamma\frac{d}{dt}\Big(\frac{\rho_\mathrm{MDW}^\mathrm{(L)}U^\alpha}{\gamma^2}\Big).
\label{eq:eulerlagrangematleft}
\end{align}
In the second equality, we have applied the product rule of differentiation for the first term.

Correspondingly, the right hand side of Eq.~\eqref{eq:eulerlagrangematL} is rewritten as
\begin{align}
 &-\gamma\frac{d}{dt}\Big(\frac{\partial\mathcal{L}_\mathrm{field}}{\partial U_\alpha}\Big)\nonumber\\
 &=-\frac{d}{dt}\Big(\gamma\frac{\partial\mathcal{L}_\mathrm{field}}{\partial U_\alpha}\Big)+\frac{d\gamma}{dt}\frac{\partial\mathcal{L}_\mathrm{field}}{\partial U_\alpha}
\nonumber\\
 &=-\frac{\partial}{\partial t}\Big(\gamma\frac{\partial\mathcal{L}_\mathrm{field}}{\partial U_\alpha}\Big)
 -\mathbf{v}_\mathrm{a}\cdot\nabla\Big(\gamma\frac{\partial\mathcal{L}_\mathrm{field}}{\partial U_\alpha}\Big)+\frac{d\gamma}{dt}\frac{\partial\mathcal{L}_\mathrm{field}}{\partial U_\alpha}.
\label{eq:eulerlagrangematright}
\end{align}
In the first equality, we have applied the product rule of differentiation, and in the second equality, we have used the material derivative obtained via the multivariate chain rule of the first term.

We next replace the left hand side the of the dynamical equation of the medium in Eq.~\eqref{eq:eulerlagrangematL} by the expression on the last row of Eq.~\eqref{eq:eulerlagrangematleft}. Correspondingly, we replace the right hand side of Eq.~\eqref{eq:eulerlagrangematL} by the last row of Eq.~\eqref{eq:eulerlagrangematright}. Thus, we obtain
\begin{align}
 &\gamma^2\rho_\mathrm{a}\frac{d^2X^\alpha}{dt^2}
+U^\alpha\frac{d(\gamma\rho_\mathrm{a})}{dt}
-\gamma\frac{d}{dt}\Big(\frac{\rho_\mathrm{MDW}^\mathrm{(L)}U^\alpha}{\gamma^2}\Big)\nonumber\\
&=-\frac{\partial}{\partial t}\Big(\gamma\frac{\partial\mathcal{L}_\mathrm{field}}{\partial U_\alpha}\Big)
 -\mathbf{v}_\mathrm{a}\cdot\nabla\Big(\gamma\frac{\partial\mathcal{L}_\mathrm{field}}{\partial U_\alpha}\Big)+\frac{d\gamma}{dt}\frac{\partial\mathcal{L}_\mathrm{field}}{\partial U_\alpha}.
 \label{eq:beforecancellation}
\end{align}

One can show that the sums of the second and third terms on the left and right hand sides of Eq.~\eqref{eq:beforecancellation} are equal in the special case of the L frame, i.e.,
\begin{align}
 & U^\alpha\frac{d(\gamma\rho_\mathrm{a})}{dt}
-\gamma\frac{d}{dt}\Big(\frac{\rho_\mathrm{MDW}^\mathrm{(L)}U^\alpha}{\gamma^2}\Big)\nonumber\\
 &=-\mathbf{v}_\mathrm{a}\cdot\nabla\Big(\gamma\frac{\partial\mathcal{L}_\mathrm{field}}{\partial U_\alpha}\Big)+\frac{d\gamma}{dt}\frac{\partial\mathcal{L}_\mathrm{field}}{\partial U_\alpha}.
 \label{eq:cancellation}
\end{align}
This result is obtained as follows. By integrating both sides of Eq.~\eqref{eq:eulerlagrangemat} with respect to proper time and denoting the integration constants by $C^\alpha$, we obtain
\begin{equation}
 \rho_0U^\alpha+\frac{\partial\mathcal{L}_\mathrm{field}}{\partial U_\alpha}=C^\alpha.
 \label{eq:Lframecondition}
\end{equation}
In the case of the L frame, where all atomic velocities result from the optical force, $\partial\mathcal{L}_\mathrm{field}/\partial U_\alpha=0$ for $\alpha=1,2,3$ in the particular point of the space-time where the fields are zero. This can be concluded from Eq.~\eqref{eq:derivativerelations1} since the field tensors that are made of field components are zero at this particular point. In this point, the atomic velocity must also be zero by the definition of the L frame and we have $\rho_0U^\alpha=0$ for $\alpha=1,2,3$. Thus, we must have $C^\alpha=0$ for $\alpha=1,2,3$, but since $C^\alpha$ are integration constants, they must be zero everywhere in the space-time. From Eqs.~\eqref{eq:derivativerelations1} and \eqref{eq:Lframecondition} and the comparison of $\rho_0 U^\alpha$ and $\partial\mathcal{L}_\mathrm{field}/\partial U_\alpha$ for $\alpha=1,2,3$, it then follows that
\begin{equation}
 \rho_0=\frac{1}{\gamma c}\frac{|(F_{\;\,\mu}^{i}\mathcal{D}_{\;\,0}^\mu-\mathcal{D}_{\;\,\mu}^{i} F_{\;\,0}^\mu)\mathbf{e}_i|}{|U^j\mathbf{e}_j|}.
\end{equation}
Thus, Eq.~\eqref{eq:derivativerelations1} gives $\partial\mathcal{L}_\mathrm{field}/\partial U_0=-\rho_0U^0$, and in Eq.~\eqref{eq:Lframecondition} for $\alpha=0$, we must have $C^0=0$. We then have $C^\alpha=0$ for all $\alpha=0,1,2,3$ at every point of the space-time.

Therefore, in the L frame, we have $\partial\mathcal{L}_\mathrm{field}/\partial U_\alpha=-\rho_0U^\alpha$. Using this equation and the definition of the mass densities $\rho_\mathrm{a}$ and $\rho_\mathrm{MDW}^\mathrm{(L)}$, given in Eqs.~\eqref{eq:mprelation1} and \eqref{eq:mprelation2}, it is then a technical but straightforward task to show that Eq.~\eqref{eq:cancellation} is satisfied identically. Using Eq.~\eqref{eq:cancellation}, the expression of the dynamical equation of the medium in the L frame in Eq.~\eqref{eq:beforecancellation} can then be written as
\begin{equation}
 \gamma^2\rho_\mathrm{a}\frac{d^2X^\alpha}{dt^2}=-\frac{\partial}{\partial t}\Big(\gamma\frac{\partial\mathcal{L}_\mathrm{field}}{\partial U_\alpha}\Big).
 \label{eq:eulerian}
\end{equation}
This is the Eulerian form of the Lagrangian dynamical equation of the medium, given in Eq.~\eqref{eq:eulerlagrangematL}. In the Eulerian form of the optical force on the right hand side of Eq.~\eqref{eq:eulerian}, there is a partial time derivative in contrast to the total time derivative in the Lagrangian form in Eq.~\eqref{eq:eulerlagrangematL}. This reflects the different views to the flow velocity field of the medium between the Lagrangian and Eulerian approaches.

In the OCD simulations of our previous works \cite{Partanen2017c,Partanen2018a,Partanen2018b}, we have solved the dynamical equation of the medium in Eq.~\eqref{eq:eulerian} in the limit of small atomic velocities in the L frame. In this approximation, we set $\gamma=1$, and also include the elastic forces between the atoms. In the present work, we neglect the elastic forces, and thus, the elastic force density is not present in Eq.~\eqref{eq:eulerian}.

To derive the SEM tensor of the medium, we write the left hand side of Eq.~\eqref{eq:eulerian} as
\begin{align}
 \gamma^2\rho_\mathrm{a}\frac{d^2X^\alpha}{dt^2}
 &=\gamma\rho_\mathrm{a}U^\beta\partial_\beta(U^\alpha/\gamma)\nonumber\\
 &=\gamma\rho_\mathrm{a}U^\beta\partial_\beta(U^\alpha/\gamma)
+(U^\alpha/\gamma)\partial_\beta(\gamma\rho_\mathrm{a}U^\beta)\nonumber\\[3pt]
&=\partial_\beta(\rho_\mathrm{a}U^\alpha U^\beta),
\label{eq:eulerlagrangematleft2}
\end{align}
In the first equality, we have used the well-known identity $\gamma d/dt=d/d\tau=U^\beta\partial_\beta$. In the second equality, we have added the term $(U^\alpha/\gamma)\partial_\beta(\gamma\rho_\mathrm{a}U^\beta)$ that is equal to zero due to the mass continuity equation, given by $\partial_\beta(\gamma\rho_\mathrm{a}U^\beta)=0$. In the final step, we have applied the product rule of derivatives to combine the two terms.

Thus, the Eulerian dynamical equations of the medium in the L frame, given in Eq.~\eqref{eq:eulerian}, are rewritten as
\begin{equation}
 \partial_\beta(\rho_aU^\alpha U^\beta)=-\frac{\partial}{\partial t}\Big(\gamma\frac{\partial\mathcal{L}_\mathrm{field}}{\partial U_\alpha}\Big).
 \label{eq:eulerianconserv}
\end{equation}
Using the definition of the SEM tensor in Eq.~\eqref{eq:emt}, we can identify, on the left hand side of Eq.~\eqref{eq:eulerianconserv}, the SEM tensor of the medium $(T_\mathrm{mat}^\mathrm{(L)})^{\alpha\beta}$, which has a particularly simple symmetric form in the L frame, as
\begin{equation}
 (T_\mathrm{mat}^\mathrm{(L)})^{\alpha\beta}=\rho_aU^\alpha U^\beta.
 \label{eq:matsem}
\end{equation}
The SEM tensor of the medium is symmetric in the L frame as it is fully determined by the mass density and the velocity components resulting from the optical force. The generally asymmetric form of the medium SEM tensor in the G frame is presented in Sec.~\ref{sec:gframe}.

In Eq.~\eqref{eq:eulerianconserv}, the four-divergence of this SEM tensor of the medium is equal to the Eulerian form of the optical force density $(f_\mathrm{opt}^\mathrm{(L)})^\alpha$, given by
\begin{align}
 (f_\mathrm{opt}^\mathrm{(L)})^\alpha
 &=-\frac{\partial}{\partial t}\Big(\gamma\frac{\partial\mathcal{L}_\mathrm{field}}{\partial U_\alpha}\Big)\nonumber\\
 &=\frac{\partial}{c\partial t}(F_{\;\,\mu}^{\alpha}\mathcal{D}_{\;\,0}^\mu-\mathcal{D}_{\;\,\mu}^{\alpha} F_{\;\,0}^\mu)\nonumber\\
 &=\frac{\partial}{\partial t}\Big(\mathbf{D}\times\mathbf{B}-\frac{\mathbf{E}\times\mathbf{H}}{c^2}\Big),
 \label{eq:opticalforce}
\end{align}
where on the last row we have the well-known expression of the Abraham force.

The dynamical equation of the medium in Eq.~\eqref{eq:eulerianconserv} can be written in a compact form as
\begin{equation}
 \partial_\beta(T_\mathrm{mat}^\mathrm{(L)})^{\alpha\beta}=(f_\mathrm{opt}^\mathrm{(L)})^\alpha.
 \label{eq:euleriancompact}
\end{equation}
Thus, the Abraham force follows \emph{ab initio} from the Lagrangian formulation of the field and the medium dynamics in the MP theory of light. In previous optics literature, the Abraham force is obtained as an \emph{ad-hoc} assumption \cite{Obukhov2008,Ramos2015} or its derivation has at least required introduction of additional arguments \cite{Gordon1973,Milonni2010}. The Abraham force is a fundamental element of the physically consistent dynamical coupling of the field and the medium and it transforms in a form-invariant way between inertial frames as detailed in Ref.~\cite{Partanen2019a}.

\subsection{\label{sec:electromagneticemtensor}SEM tensor of the electromagnetic field}

Next, we derive the SEM tensor of the electromagnetic field in the medium. The beginning of this section follows the conventional approach presented, e.g., in the well-known textbook of Landau and Lifshitz in Ref.~\cite{Landau1989} for the fields in vacuum. For light in the medium, the relation between the tensors $\mathcal{D}^{\alpha\beta}$ and $F^{\alpha\beta}$ as given in Eq.~\eqref{eq:displacementtensorfourpotential} is slightly different from that in vacuum, but this difference does not influence the first part of our derivation below. Our first goal is to write a divergenceless tensor that is made of second-order electric and magnetic field quantities and that coincides the standard definition of the SEM tensor of the electromagnatic field in the case of vacuum. Second, by accounting for the optical force in Eq.~\eqref{eq:opticalforce}, we will write the SEM tensor of the electromagnetic field for which the four-divergence is equal to the optical force, as required by the full consistency with the dynamical equation of the medium in Eq.~\eqref{eq:euleriancompact}.

To derive a divergenceless tensor made of second-order electric and magnetic field quantities, we first express $\partial_\alpha\mathcal{L}_\mathrm{field}$ by using the chain rule as
\begin{equation}
 \partial_\alpha\mathcal{L}_\mathrm{field}=\frac{\partial\mathcal{L}_\mathrm{field}}{\partial A_\mu}\partial_\alpha A_\mu
+\frac{\partial\mathcal{L}_\mathrm{field}}{\partial(\partial_\beta A_\mu)}\partial_\alpha(\partial_\beta A_\mu).
\label{eq:chainrule2}
\end{equation}
Substituting the Euler-Lagrange equation in Eq.~\eqref{eq:EulerLagrange1} into the factor of the first term in Eq.~\eqref{eq:chainrule2}, using $\partial_\alpha\partial_\beta A_\mu=\partial_\beta\partial_\alpha A_\mu$ in the second term, and using $\partial_\alpha\mathcal{L}_\mathrm{field}=\delta_\alpha^\beta\partial_\beta\mathcal{L}_\mathrm{field}$ on the left hand side of Eq.~\eqref{eq:chainrule2} gives
\begin{align}
 \delta_\alpha^\beta\partial_\beta\mathcal{L}_\mathrm{field} &=\partial_\beta\Big[\frac{\partial\mathcal{L}_\mathrm{field}}{\partial(\partial_\beta A_\mu)}\Big]\partial_\alpha A_\mu
+\frac{\partial\mathcal{L}_\mathrm{field}}{\partial(\partial_\beta A_\mu)}\partial_\beta(\partial_\alpha A_\mu)\nonumber\\
&=\partial_\beta\Big[\frac{\partial\mathcal{L}_\mathrm{field}}{\partial(\partial_\beta A_\mu)}\partial_\alpha A_\mu\Big].
\label{eq:emsemderivation1}
\end{align}
Using Eq.~\eqref{eq:EulerLagrangeParts} on the right hand side of Eq.~\eqref{eq:emsemderivation1}, moving both terms of Eq.~\eqref{eq:emsemderivation1} on the same side, and using the expression of the Lagrangian density of the field in Eq.~\eqref{eq:LagrangianDensityField}, then gives
\begin{align}
&\partial_\beta\Big[\frac{\partial\mathcal{L}_\mathrm{field}}{\partial(\partial_\beta A_\mu)}\partial_\alpha A_\mu-\delta_\alpha^\beta\mathcal{L}_\mathrm{field}\Big]\nonumber\\
&=\partial_\beta\Big[-\mathcal{D}^{\beta\mu}\partial_\alpha A_\mu+\frac{1}{4}\delta_\alpha^\beta F_{\mu\nu}\mathcal{D}^{\mu\nu}\Big]=0.
 \label{eq:emsemderivation2}
\end{align}

It is well-known that adding a tensor of the form $\partial_\mu\Psi^{\alpha\beta\mu}$, where $\Psi^{\alpha\beta\mu}$ satisfies $\Psi^{\alpha\beta\mu}=-\Psi^{\alpha\mu\beta}$, to any tensor does not change the value of the four-divergence since we have an identity $\partial_\beta\partial_\mu\Psi^{\alpha\beta\mu}=0$ \cite{Landau1989}. In the same way as made in the case of vacuum in Ref.~\cite{Landau1989}, to make the tensor, from which the four-divergence is taken in Eq.~\eqref{eq:emsemderivation2}, gauge invariant, we add $\partial_\mu\Psi_\alpha^{\;\,\beta\mu}=\mathcal{D}^{\beta\mu}\partial_\mu A_\alpha$ under the four-divergence in Eq.~\eqref{eq:emsemderivation2}. Applying the Minkowski metric tensor $g^{\alpha\nu}$ to rise the index $\alpha$, we can then write Eq.~\eqref{eq:emsemderivation2} as
\begin{equation}
 \partial_\beta\Big[F_{\;\,\mu}^\alpha\mathcal{D}^{\mu\beta}+\frac{1}{4}g^{\alpha\beta} F_{\mu\nu}\mathcal{D}^{\mu\nu}\Big]=0.
\label{eq:emsemderivation3}
\end{equation}
In Eq.~\eqref{eq:emsemderivation3}, we can identify a divergenceless tensor that is made of second-order electric and magnetic field quantities and that coincides with the standard definition of the SEM tensor of the electromagnetic field in the case of vacuum. Thus, we have now achieved the goal of the first part of our derivation.

We are now ready to introduce the SEM tensor $(T_\mathrm{field}^\mathrm{(L)})^{\alpha\beta}$ of the electromagnetic field in the MP theory of light. To be in accordance with the law of action and counteraction and consistent with the dynamical equation of the medium in Eq.~\eqref{eq:euleriancompact}, this tensor must satisfy \cite{Partanen2019a}
\begin{equation}
 \partial_\beta(T_\mathrm{field}^\mathrm{(L)})^{\alpha\beta}=-\partial_\beta(T_\mathrm{mat}^\mathrm{(L)})^{\alpha\beta}=-(f_\mathrm{opt}^\mathrm{(L)})^\alpha.
 \label{eq:actionreaction1}
\end{equation}
The optical force in Eq.~\eqref{eq:opticalforce} can be written as
$(f_\mathrm{opt}^\mathrm{(L)})^\alpha=\partial_\beta[\delta_0^\beta(F_{\;\,\mu}^{\alpha}\mathcal{D}_{\;\,0}^\mu-\mathcal{D}_{\;\,\mu}^{\alpha} F_{\;\,0}^\mu)]$. By subtracting this expression of the optical force from both sides of Eq.~\eqref{eq:emsemderivation3} and by comparing the resulting equation with Eq.~\eqref{eq:actionreaction1}, we can then identify the expression of the SEM tensor of the electromagnetic field, given by
\begin{align}
 &(T_\mathrm{field}^\mathrm{(L)})^{\alpha\beta}\nonumber\\
 &=F_{\;\,\mu}^\alpha\mathcal{D}^{\mu\beta}+\frac{1}{4}g^{\alpha\beta} F_{\mu\nu}\mathcal{D}^{\mu\nu}-\delta_0^\beta(F_{\;\,\mu}^{\alpha}\mathcal{D}_{\;\,0}^\mu-\mathcal{D}_{\;\,\mu}^{\alpha} F_{\;\,0}^\mu).
 \label{eq:emsem}
\end{align}
This SEM tensor of the field is found to be equal to the conventional Abraham SEM tensor. In some previous works \cite{Obukhov2008,Ramos2015}, it has been argued that this form of the Abraham SEM tensor would only be valid in the L frame. However, in the MP theory of light, the SEM tensor in Eq.~\eqref{eq:emsem} has been shown to be a valid SEM tensor of the electromagnetic field part of the coupled state of the field and the medium in an arbitrary inertial frame \cite{Partanen2019a}. Also, note that the tensor under the four-divergence in Eq.~\eqref{eq:emsemderivation3} is the conventional Minkowski SEM tensor that has been discussed in some depth in our previous work \cite{Partanen2019a}.

\subsection{Total SEM tensor of the field and the medium}

In previous sections, we derived the electromagnetic field and the medium parts of the total SEM tensor of the system. Here, we summarize how these SEM tensor parts constitute the total SEM tensor, which obeys all conservation laws of energy, momentum, and angular momentum. The total SEM tensor of the electromagnetic field and the medium is given by the sum of the field and the medium parts as
\begin{equation}
 (T_\mathrm{tot}^\mathrm{(L)})^{\alpha\beta}=(T_\mathrm{field}^\mathrm{(L)})^{\alpha\beta}+(T_\mathrm{mat}^\mathrm{(L)})^{\alpha\beta}.
 \label{eq:totsem}
\end{equation}

The optical force density appears with different signs in the dynamical equations of the field and the medium as dictated by the law of action and counteraction in Eq.~\eqref{eq:actionreaction1}. Therefore, the four-divergence of the total SEM tensor in Eq.~\eqref{eq:totsem} is zero as
\begin{equation}
 \partial_\beta(T_\mathrm{tot}^\mathrm{(L)})^{\alpha\beta}=0.
 \label{eq:totsemdivergence}
\end{equation}
The total SEM tensor in Eq.~\eqref{eq:totsem} is also symmetric. Thus, it fulfills all the conservation laws of energy, momentum, and angular momentum. Consequently, this tensor is the total conserved Poincar\'e current of the field and the medium \cite{Bliokh2013b}. This is a strong argument for the ultimate consistency of the MP theory of light.

\subsection{SEM tensor of the coupled MP state of light}

In our previous work \cite{Partanen2019a}, in addition to the SEM tensor of the field, we investigated in detail only the part of the medium SEM tensor associated with the atomic MDW. By definition, the atomic MDW includes only the dynamics of atoms which represents the difference between the true and the equilibrium atomic densities of the medium. Thus, the SEM tensor of the atomic MDW is only a small part of the total SEM tensor of the medium and it is equal to the deviation of the total SEM tensor of the medium from its equilibrium value. The SEM tensor of the atomic MDW in the MP theory of light is given in the L frame by \cite{Partanen2019a}
\begin{equation}
 (T_\mathrm{MDW}^\mathrm{(L)})^{\alpha\beta}=(T_\mathrm{mat}^\mathrm{(L)})^{\alpha\beta}-(T_\mathrm{mat,0}^\mathrm{(L)})^{\alpha\beta},
 \label{eq:mdwsem}
\end{equation}
where $(T_\mathrm{mat,0}^\mathrm{(L)})^{\alpha\beta}$ is the SEM tensor of the medium in the L frame in the absence of the electromagnetic field. Thus, this tensor is given in terms of the unperturbed rest mass density $\rho_0^\mathrm{(L)}$ of the medium as $(T_\mathrm{mat,0}^\mathrm{(L)})^{\alpha\beta}=\rho_0^\mathrm{(L)}c^2\delta_0^\alpha\delta_0^\beta$.

The total SEM tensor of the coupled MP state of light is given by the sum of the SEM tensors of the electromagnetic field and the atomic MDW as \cite{Partanen2019a}
\begin{equation}
 (T_\mathrm{MP}^\mathrm{(L)})^{\alpha\beta}=(T_\mathrm{field}^\mathrm{(L)})^{\alpha\beta}+(T_\mathrm{MDW}^\mathrm{(L)})^{\alpha\beta}.
 \label{eq:mpsem}
\end{equation}

Since the four-divergence of the second term in Eq.~\eqref{eq:mdwsem} is zero, it immediately follows from Eqs.~\eqref{eq:totsem} and \eqref{eq:totsemdivergence} that the four-divergence of the MP SEM tensor in Eq.~\eqref{eq:mpsem} is also zero as
\begin{equation}
 \partial_\beta(T_\mathrm{MP}^\mathrm{(L)})^{\alpha\beta}=0.
 \label{eq:mpsemdivergence}
\end{equation}
Since the MP SEM tensor in Eq.~\eqref{eq:mpsem} is also symmetric, this tensor is the total conserved Poincar\'e current of light. This is a strong argument for the MP SEM tensor to be the unique physically correct SEM tensor of light. The Lorentz covariance of the MP theory of light and the related transformation of the SEM tensors between arbitrary inertial frames are described in Ref.~\cite{Partanen2019a}. The Lorentz transformations of the fields and the atomic MDW quantities are also briefly reviewed in Appendix \ref{apx:Lorentz}.

\section{\label{sec:gframe}SEM tensors in the G frame}
In previous sections, we have derived the SEM tensors of the medium and the electromagnetic field in the special case of the L frame. In this section, we generalize the results for the G frame. Using the Lorentz transformations of the fields and the atomic MDW quantities, given in Appendix \ref{apx:Lorentz}, the SEM tensors can be transformed from the L frame into an arbitrary inertial frame. Consequently, the SEM tensor of the atomic MDW is given in the G frame by
\begin{equation}
 T_\mathrm{MDW}^{\alpha\beta}=\rho_\mathrm{MDW}^\mathrm{(G)}V_\mathrm{a}^\alpha V_\mathrm{l}^\beta
 +\rho_\mathrm{MDW}^\mathrm{(G)}c\delta_0^\beta(V_\mathrm{l}^\alpha-V_\mathrm{a}^\alpha),
\label{eq:mdwsemG}
\end{equation}
where we define $V_\mathrm{a}^\alpha=U^\alpha/\gamma=(c,v_\mathrm{a}^x,v_\mathrm{a}^y,v_\mathrm{a}^z)$ and $V_\mathrm{l}^\alpha=(c,v_\mathrm{l}^x,v_\mathrm{l}^y,v_\mathrm{l}^z)$, in which $v_\mathrm{l}^x$, $v_\mathrm{l}^y$, and $v_\mathrm{l}^z$ are components of the velocity of light. These velocity components of light can also be obtained from the electric and magnetic fields as given in Eq.~\eqref{eq:velocityoflight} of Appendix \ref{apx:materialparameters}.

The SEM tensor of the equilibrium mass density of the medium is well known to be given by
\begin{equation}
 T_\mathrm{mat,0}^{\alpha\beta}=\gamma_\mathrm{rel}^2\rho_0^\mathrm{(L)}V_\mathrm{rel}^\alpha V_\mathrm{rel}^\beta,
\label{eq:matsem0G}
\end{equation}
where $V_\mathrm{rel}^\alpha=(c,v_\mathrm{rel}^x,v_\mathrm{rel}^y,v_\mathrm{rel}^z)$, in which $v_\mathrm{rel}^x$, $v_\mathrm{rel}^y$, and $v_\mathrm{rel}^z$ are components of the relative velocity of the L frame with respect to the G frame, and $\gamma_\mathrm{rel}$ is the corresponding Lorentz factor.

In agreement with Eq.~\eqref{eq:mdwsem}, the total SEM tensor of the medium is the sum of the SEM tensors in Eqs.~\eqref{eq:mdwsemG} and Eq.~\eqref{eq:matsem0G} as
\begin{align}
 T_\mathrm{mat}^{\alpha\beta} &=\gamma_\mathrm{rel}^2\rho_0^\mathrm{(L)}V_\mathrm{rel}^\alpha V_\mathrm{rel}^\beta
+\rho_\mathrm{MDW}^\mathrm{(G)}V_\mathrm{a}^\alpha V_\mathrm{l}^\beta\nonumber\\
&\hspace{0.4cm}+\rho_\mathrm{MDW}^\mathrm{(G)}c\delta_0^\beta(V_\mathrm{l}^\alpha-V_\mathrm{a}^\alpha).
\label{eq:matsemG}
\end{align}
It is straightforward to check that in the special case of the L frame, where $V_\mathrm{rel}^\alpha=c\delta_0^\alpha$, the SEM tensor of the medium in Eq.~\eqref{eq:matsemG} is equal to the SEM tensor in Eq.~\eqref{eq:matsem}.

In agreement with our previous work \cite{Partanen2019a}, the SEM tensor of the electromagnetic field in the G frame is of the same form as in the L-frame in Eq.~\eqref{eq:emsem}. Therefore, it is given by
\begin{equation}
 T_\mathrm{field}^{\alpha\beta}
=F_{\;\,\mu}^\alpha\mathcal{D}^{\mu\beta}+\frac{1}{4}g^{\alpha\beta} F_{\mu\nu}\mathcal{D}^{\mu\nu}-\delta_0^\beta(F_{\;\,\mu}^{\alpha}\mathcal{D}_{\;\,0}^\mu-\mathcal{D}_{\;\,\mu}^{\alpha} F_{\;\,0}^\mu).
 \label{eq:emsemG}
\end{equation}

The SEM tensor of the MP state of light is the sum of the SEM tensors in Eqs.~\eqref{eq:mdwsemG} and \eqref{eq:emsemG} and it becomes
\begin{align}
 T_\mathrm{MP}^{\alpha\beta}
&=F_{\;\,\mu}^\alpha\mathcal{D}^{\mu\beta}+\frac{1}{4}g^{\alpha\beta} F_{\mu\nu}\mathcal{D}^{\mu\nu}-\delta_0^\beta(F_{\;\,\mu}^{\alpha}\mathcal{D}_{\;\,0}^\mu-\mathcal{D}_{\;\,\mu}^{\alpha} F_{\;\,0}^\mu)\nonumber\\
&\hspace{0.5cm}+\rho_\mathrm{MDW}^\mathrm{(G)}V_\mathrm{a}^\alpha V_\mathrm{l}^\beta
+\rho_\mathrm{MDW}^\mathrm{(G)}c\delta_0^\beta(V_\mathrm{l}^\alpha-V_\mathrm{a}^\alpha)\nonumber\\
&=F_{\;\,\mu}^\alpha\mathcal{D}^{\mu\beta}+\frac{1}{4}g^{\alpha\beta} F_{\mu\nu}\mathcal{D}^{\mu\nu}
+\rho_\mathrm{MDW}^\mathrm{(G)}V_\mathrm{a}^\alpha V_\mathrm{l}^\beta,
 \label{eq:mpsemG}
\end{align}
where the third and fifth terms of the first expression cancel each other. 
Correspondingly, the total SEM tensor of the field and the medium is the sum of the SEM tensors in Eqs.~\eqref{eq:matsemG} and \eqref{eq:emsemG} and it becomes
\begin{align}
 T_\mathrm{tot}^{\alpha\beta}
&=F_{\;\,\mu}^\alpha\mathcal{D}^{\mu\beta}+\frac{1}{4}g^{\alpha\beta} F_{\mu\nu}\mathcal{D}^{\mu\nu}
%-\delta_0^\beta(F_{\;\,\mu}^{\alpha}\mathcal{D}_{\;\,0}^\mu-\mathcal{D}_{\;\,\mu}^{\alpha} F_{\;\,0}^\mu)\nonumber\\
+\rho_\mathrm{MDW}^\mathrm{(G)}V_\mathrm{a}^\alpha V_\mathrm{l}^\beta\nonumber\\
&\hspace{0.5cm}+\gamma_\mathrm{rel}^2\rho_0^\mathrm{(L)}V_\mathrm{rel}^\alpha V_\mathrm{rel}^\beta.
 \label{eq:totsemG}
\end{align}
The only difference between the MP SEM tensor in Eq.~\eqref{eq:mpsemG} and the total SEM tensor in Eq.~\eqref{eq:totsemG} is that the MP SEM tensor excludes the contribution of the equilibrium mass density of the medium, which is described by the SEM tensor in Eq.~\eqref{eq:matsem0G}.

The SEM tensors of this section are all form-invariant between arbitrary inertial frames, i.e., these tensors in all inertial frames are formed in the same way from the fields, light and atomic velocities, the space and time coordinates, and the mass densities that transform according to the Lorentz transformation as described in Appendix \ref{apx:Lorentz}. Note that only Lorentz-covariant second-rank tensors transform between inertial frames according to the matrix equation $(T')^{\alpha\beta}=\Lambda_{\;\,\mu}^\alpha T^{\mu\nu}\Lambda_{\;\,\nu}^\beta$. Thus, the SEM tensor of the atomic MDW in Eq.~\eqref{eq:mdwsemG} and the SEM tensor of the electromagnetic field in Eq.~\eqref{eq:emsemG} do not separately satisfy this equation, but this equation is satisfied for their sum, which is the total MP SEM tensor of light, given in Eq.~\eqref{eq:mpsemG}. The same applies to the field and medium parts of the total SEM tensor of the system in Eqs.~\eqref{eq:matsemG} and \eqref{eq:emsemG} and their sum in Eq.~\eqref{eq:totsemG}. In the present work, other Lorentz-covariant second-rank tensors for which the matrix equation above is satisfied include the electromagnetic field and displacement tensors $F^{\alpha\beta}$ and $\mathcal{D}^{\alpha\beta}$ in Eqs.~\eqref{eq:electromagnetictensor} and \eqref{eq:displacementtensor} and the SEM tensor of the equilibrium mass density of the medium in Eq.~\eqref{eq:matsem0G}.

The zero four-divergence of the SEM tensor is directly related to the conservation laws \cite{Jackson1999,Landau1989}, and thus, the SEM tensors of interacting subsystems, such as the electromagnetic field and the atomic MDW, do not separately satisfy this condition. This conclusion and the Lorentz covariance of the MP SEM tensor are in full agreement with the results of our previous work \cite{Partanen2019a}.

\section{\label{sec:solution}Exact solution of the coupled dynamical equations}

In this section, we demonstrate the simultaneous solution of the dynamical equations of the field and the medium by considering a light pulse in the L frame. The following example Gaussian light pulse, including both the field and the associated atomic MDW, was found to simultaneously fulfill the Euler-Lagrange equations of the field and the medium, given in Eqs.~\eqref{eq:eulerlagrangemat1} and \eqref{eq:EulerLagrange1}. The pertinent Euler-Lagrange equations lead to the Gauss-Ampere law in Eq.~\eqref{eq:EulerLagrangeD} for the field and to the Newtonian equation of the medium, given in Eq.~\eqref{eq:eulerian}, with the optical force, given in Eq.~\eqref{eq:opticalforce}. Note that the form of Maxwell's equations is standard, but these equations are through the permittivity and permeability of the medium coupled to the dynamical state of the medium via the atomic velocity $\mathbf{v}_\mathrm{a}$. See the comments at the end of Sec.~\ref{sec:EulerLagrangeField}.

The simultaneous field-medium solution of the Euler-Lagrange equations was found heuristically and with some experimenting. The solution is presented through Eqs.~\eqref{eq:efield} and \eqref{eq:atomicvelocity} and the reader can directly verify that the solution fulfills the dynamical equations in Eq.~\eqref{eq:EulerLagrangeD} and \eqref{eq:eulerian}. The reader can also use the solution of the coupled state to calculate the SEM tensors for the field and the medium and the total SEM tensor of the light pulse. One can also study the covariance and the divergence properties of the SEM tensors to verify that the results are consistent with the results of our previous paper \cite{Partanen2019a}. Finally, below, we also quantitatively demonstrate the small magnitude of the kinetic energy density carried by the medium when a short 1-$\mu$J Gaussian pulse is propagating in a silicon crystal.

\subsection{Electric and magnetic fields}

As an example, we take the sinusoidal electric field $\mathbf{E}^\mathrm{(L)}(z,t)$ of a Gaussian plane wave pulse in the L frame of a nondispersive medium, which is given by
\begin{align}
 &\mathbf{E}^\mathrm{(L)}(z,t)\nonumber\\
&=\omega_0^\mathrm{(L)}\mathcal{E}\sin[k^\mathrm{(L)}(z-ct/n^\mathrm{(L)})]e^{-(\Delta k^\mathrm{(L)})^2(z-ct/n^\mathrm{(L)})^2/2}\hat{\mathbf{x}}.
\label{eq:efield}
\end{align}
In Eq.~\eqref{eq:efield}, $\omega_0^\mathrm{(L)}$ is the central angular frequency, $k^\mathrm{(L)}=n^\mathrm{(L)}\omega_0^\mathrm{(L)}/c$ is the wave vector, $\Delta k^\mathrm{(L)}=n^\mathrm{(L)}\Delta\omega_0^\mathrm{(L)}/c$ is the standard deviation of $k^\mathrm{(L)}$, $\Delta\omega_0^\mathrm{(L)}$ is the spectral width, and $\mathcal{E}$ is the normalization constant. The central angular frequency $\omega_0^\mathrm{(L)}$ of the L frame is related to the angular frequency $\omega_0^\mathrm{(A)}$ in the A frame by the well-known relativistic Doppler shift, $\omega_0^\mathrm{(A)}=\gamma(1-n^\mathrm{(L)}v_\mathrm{a}/c)\omega_0^\mathrm{(L)}$. The refractive index $n^\mathrm{(L)}$ of the L frame is related to the proper refractive index $n^\mathrm{(A)}$, defined in the A frame, by the relativistic velocity addition formula as described in Appendix \ref{apx:materialparameters}.

In the following, we keep the expression of the electric field in Eq.~\eqref{eq:efield} as fixed and derive analytic expressions for the fields $\mathbf{B}^\mathrm{(L)}$, $\mathbf{D}^\mathrm{(L)}$, and $\mathbf{H}^\mathrm{(L)}$ and the dynamical variables of the medium. The magnetic flux density $\mathbf{B}^\mathrm{(L)}=n^\mathrm{(L)}\hat{\mathbf{z}}\times\mathbf{E}^\mathrm{(L)}/c$ follows directly from Faraday's law. Once we know the electric field $\mathbf{E}^\mathrm{(L)}$ and the magnetic flux density $\mathbf{B}^\mathrm{(L)}$, we can express the electromagnetic field tensor $F^{\alpha\beta}$ in terms of these fields as presented in Eq.~\eqref{eq:electromagnetictensor}. After this, we can write the electromagnetic displacement tensor $\mathcal{D}^{\alpha\beta}$ by using Eq.~\eqref{eq:displacementtensorfourpotential}, where the atomic velocity $\mathbf{v}_\mathrm{a}^\mathrm{(L)}$ in the matrix $h^{\alpha\beta}$ is unknown at this stage. We find that the fields $\mathbf{D}^\mathrm{(L)}$ and $\mathbf{H}^\mathrm{(L)}$ are given in terms of the fields $\mathbf{E}^\mathrm{(L)}$ and $\mathbf{B}^\mathrm{(L)}$ as $\mathbf{D}^\mathrm{(L)}=\varepsilon^\mathrm{(L)}\mathbf{E}^\mathrm{(L)}$ and $\mathbf{B}^\mathrm{(L)}=\mu^\mathrm{(L)}\mathbf{H}^\mathrm{(L)}$, where the permittivity $\varepsilon^\mathrm{(L)}$ and the permeability $\mu^\mathrm{(L)}$ of the L frame are related to the proper permittivity and permeability as detailed in Appendix \ref{apx:materialparameters}. If the material is exactly nondispersive, $n^\mathrm{(L)}$ must be constant so that all frequency components propagate with equal velocity $\mathbf{v}_\mathrm{l}^\mathrm{(L)}=(c/n^\mathrm{(L)})\hat{\mathbf{z}}$. Correspondingly, in this section, we assume that $\varepsilon^\mathrm{(L)}$ and $\mu^\mathrm{(L)}$ are constants.

\subsection{Atomic velocity}

The fields $\mathbf{E}^\mathrm{(L)}$, $\mathbf{B}^\mathrm{(L)}$, $\mathbf{D}^\mathrm{(L)}$, and $\mathbf{H}^\mathrm{(L)}$ as presented above automatically satisfy all Maxwell's equations independently of the atomic velocity. This follows from the relations between the fields above, in which the atomic velocity is only implicitly present through the transformations of the material parameters $\varepsilon^\mathrm{(L)}$, $\mu^\mathrm{(L)}$, and $n^\mathrm{(L)}$. In spite of this, the atomic velocity in the L frame becomes unambiguously determined by Newton's equation of motion for the medium that depends on the optical force as described in Secs.~\ref{sec:EulerLagrangeMedium} and \ref{sec:SEMtensorMedium}. In the L frame, the atomic velocity has no components independent of the optical force. Thus, the atomic velocity can be uniquely solved by using Eq.~\eqref{eq:Lframecondition}, where the right hand side is zero for the L frame.

By taking the expression of the derivative of the Lagrangian density of the field in Eq.~\eqref{eq:derivativerelations1} and expressing the four-velocity and the Lorentz factor in terms of the atomic velocity $\mathbf{v}_\mathrm{a}^\mathrm{(L)}$, we find that the atomic velocity is given by
\begin{equation}
 \mathbf{v}_\mathrm{a}^\mathrm{(L)}=\Big(n^\mathrm{(L)}-\frac{1}{n^\mathrm{(L)}}\Big)\frac{\varepsilon^\mathrm{(L)}|\mathbf{E}^\mathrm{(L)}|^2}{\rho_0^\mathrm{(L)}c}\hat{\mathbf{z}}.
\label{eq:atomicvelocity}
\end{equation}
The time component of Eq.~\eqref{eq:Lframecondition} is satisfied identically with the velocity components in Eq.~\eqref{eq:atomicvelocity}, which were obtained from the space components of Eq.~\eqref{eq:Lframecondition}. This is as expected since the three-velocity components determine both the time and space components of the four-velocity.

\subsection{Disturbed mass density of the medium}

The true disturbed atomic rest mass density $\rho_\mathrm{a}$ and the mass density $\rho_\mathrm{MDW}^\mathrm{(L)}$ of the atomic MDW can be uniquely solved from the energy and momentum density relations of the SEM tensors of the medium and the atomic MDW. Based on Eq.~\eqref{eq:mdwsem}, we know that the total energy density of the atomic MDW is given by
\begin{equation}
 W_\mathrm{MDW}^\mathrm{(L)}=\rho_\mathrm{MDW}^\mathrm{(L)}c^2=\gamma^2\rho_\mathrm{a}c^2-\rho_0^\mathrm{(L)}c^2.
\end{equation}
The corresponding total momentum density of the MDW is given by
\begin{equation}
 \mathbf{G}_\mathrm{MDW}^\mathrm{(L)}=\rho_\mathrm{MDW}^\mathrm{(L)}\mathbf{v}_\mathrm{l}^\mathrm{(L)}=\gamma^2\rho_\mathrm{a}\mathbf{v}_\mathrm{a}^\mathrm{(L)}.
\end{equation}

From these two equations, we find that the mass density of the atomic MDW and the true disturbed rest mass density of the medium are given by
\begin{equation}
 \rho_\mathrm{MDW}^\mathrm{(L)}=\frac{n^\mathrm{(A)}n^\mathrm{(L)}-1}{c^2}\varepsilon^\mathrm{(L)}|\mathbf{E}^\mathrm{(L)}|^2,
\label{eq:mdwdensity}
\end{equation}
\begin{equation}
 \rho_\mathrm{a}=\frac{\rho_0^\mathrm{(L)}+\rho_\mathrm{MDW}^\mathrm{(L)}}{\gamma^2}.
\end{equation}
Note that the proper refractive index $n^\mathrm{(A)}$ in Eq.~\eqref{eq:mdwdensity} is given in terms of the L-frame quantities as presented in Appendix \ref{apx:materialparameters}.

\subsection{Energy and momentum densities of the field and the MDW}

Energy and momentum densities of the atomic MDW corresponding to the results of the previous subsection are given in the L frame by
\begin{equation}
 W_\mathrm{MDW}^\mathrm{(L)}=(n^\mathrm{(A)}n^\mathrm{(L)}-1)\varepsilon^\mathrm{(L)}|\mathbf{E}^\mathrm{(L)}|^2,
\label{eq:mdwenergy}
\end{equation}
\begin{equation}
 \mathbf{G}_\mathrm{MDW}^\mathrm{(L)}=\Big(n^\mathrm{(A)}-\frac{1}{n^\mathrm{(L)}}\Big)\frac{\varepsilon^\mathrm{(L)}|\mathbf{E}^\mathrm{(L)}|^2}{c}\hat{\mathbf{z}}.
\label{eq:mdwmomentum}
\end{equation}
The energy and momentum densities of the electromagnetic field are given correspondingly by
\begin{equation}
 W_\mathrm{field}^\mathrm{(L)}=\varepsilon^\mathrm{(L)}|\mathbf{E}^\mathrm{(L)}|^2,
\label{eq:fieldenergy}
\end{equation}
\begin{equation}
 \mathbf{G}_\mathrm{field}^\mathrm{(L)}=\frac{\varepsilon^\mathrm{(L)}|\mathbf{E}^\mathrm{(L)}|^2}{n^\mathrm{(L)}c}\hat{\mathbf{z}}.
\label{eq:fieldmomentum}
\end{equation}
Thus, the total energy and momentum densities of the coupled MP state of the electromagnetic field and the atomic MDW are given by
\begin{equation}
 W_\mathrm{MP}^\mathrm{(L)}=n^\mathrm{(A)}n^\mathrm{(L)}\varepsilon^\mathrm{(L)}|\mathbf{E}^\mathrm{(L)}|^2,
\label{eq:mpenergy}
\end{equation}
\begin{equation}
 \mathbf{G}_\mathrm{MP}^\mathrm{(L)}=n^\mathrm{(A)}\frac{\varepsilon^\mathrm{(L)}|\mathbf{E}^\mathrm{(L)}|^2}{c}\hat{\mathbf{z}}.
\label{eq:mpmomentum}
\end{equation}

\begin{figure*}
\centering
\includegraphics[width=\textwidth]{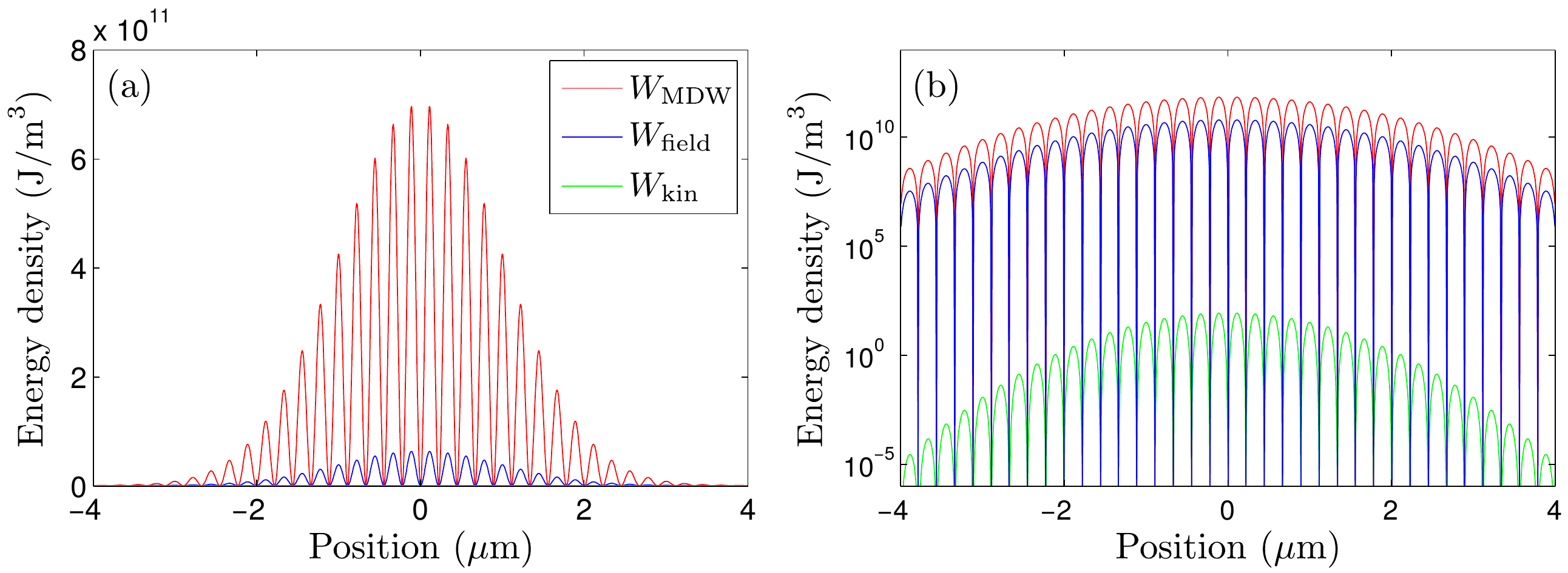}
\vspace{-0.5cm}
\caption{\label{fig:graphs}
Total energy density $W_\mathrm{MDW}$ of the atomic MDW, the energy density $W_\mathrm{field}$ of the electromagnetic field, and the kinetic energy density $W_\mathrm{kin}$ of the atomic MDW for a Gaussian plane wave pulse in silicon plotted in (a) linear and (b) logarithmic scale as a function of the position when the pulse center is at $z=0$ $\mu$m. The wavelength is $\lambda_0^\mathrm{(L)}=1550$ nm and the total electromagnetic energy of the pulse is 1 $\mu$J per circular cross-sectional area of diameter 4 $\mu$m. As shown in our previous works \cite{Partanen2017c}, the total energy density of the MP state of light is dominated by the rest energy density of the MDW. In contrast to our earlier works, the kinetic energy $W_\mathrm{kin}$ of atoms is calculated exactly in this figure. It is invisible in the linear scale (a), where the $W_\mathrm{MDW}$ line is almost entirely due to the rest energy component, but can be made visible in the logarithmic scale (b).}
\end{figure*}

\subsection{Energy densities of the field and the medium in silicon}

To illustrate our solution of the coupled dynamical equations, we plot the position dependence of the total energy density of the atomic MDW, the energy density of the electromagnetic field, and the kinetic energy density of the atomic MDW for an example Gaussian pulse using realistic material parameters of silicon. In the present illustration, we neglect the material dispersion. The plots in linear and logarithmic scales are presented in Fig.~\ref{fig:graphs}.

In our example, the angular frequency of the field is $\omega^\mathrm{(L)}=2\pi c/\lambda_0^\mathrm{(L)}$, where $\lambda_0^\mathrm{(L)}$ is the vacuum wavelength $\lambda_0^\mathrm{(L)}=1550$ nm. For this wavelength, the refractive index, permittivity, and permeability of silicon are $n^\mathrm{(L)}=3.4757$, $\varepsilon^\mathrm{(L)}=\varepsilon_0(n^\mathrm{(L)})^2$, $\mu^\mathrm{(L)}=\mu_0$ \cite{Li1980}. For the relative spectral width, we use the value $\Delta k_0^\mathrm{(L)}/k_0^\mathrm{(L)}=0.05$. The mass density of silicon is taken to be $\rho_0^\mathrm{(L)}=2329$ kg/m$^3$ \cite{Lide2004}. The normalization constant $E_0$ is determined so that the total electromagnetic energy of the pulse in the L frame is 1 $\mu$J per circular cross-sectional area of diameter 4 $\mu$m.

Figure \ref{fig:graphs}(a) shows the total energy density of the atomic MDW, the energy density of the electromagnetic field, and the kinetic energy density of the atomic MDW as a function of the position in linear scale. Under the Gaussian envelope, the functional form of the energy density of the electromagnetic field is perfectly sinusoidal as a consequence of the fact that the sinusoidal electric field in Eq.~\eqref{eq:efield} was taken as given. The energy density of the atomic MDW consists mainly of the excess rest energy of the disturbed mass density of the medium. This energy density is also closely sinusoidal. According to Eq.~\eqref{eq:mdwenergy}, the total energy density of the atomic MDW is equal to the energy density of the electromagnetic field multiplied by a factor $n^\mathrm{(A)}n^\mathrm{(L)}-1$. This factor is approximatively constant as $n^\mathrm{(A)}\approx n^\mathrm{(L)}$ applies very accurately due to the large mass energy density of silicon in comparison with the energy density of the electromagnetic field. The kinetic energy density of atoms is equal to $W_\mathrm{kin}=(\gamma-1)\gamma\rho_\mathrm{a}c^2\approx\frac{1}{2}\rho_0^\mathrm{(L)}|\mathbf{v}_\mathrm{a}^\mathrm{(L)}|^2$. This kinetic energy of atoms is not visible in the linear scale of Fig.~\ref{fig:graphs}(a) due to its extreme smallness.

Figure \ref{fig:graphs}(b) presents the quantities of Fig.~\ref{fig:graphs}(a) in the logarithmic scale to make the very small kinetic energy density of silicon atoms visible in the same graph with the larger energy density of the electromagnetic field and the mass energy density of the atomic MDW. It is seen that, in our example, the kinetic energy density of atoms is roughly eight orders of magnitude smaller than the energy density of the electromagnetic field. Thus, its effects are negligible in common laboratory experiments and photonics technologies.

\section{\label{sec:comparison}Comparison with the zero kinetic enery limit of the MP}

In our previous works, we have effectively made an approximation that the terms, which are of the second order in the velocity of atoms, are zero. In this approximation, the total momentum density of light in the L frame becomes proportional to the phase refractive index $n^\mathrm{(L)}$ as described in Ref.~\cite{Partanen2017c}. In contrast, in the present work, where the small kinetic energy terms of atoms have been accounted for, the total momentum density of light is proportional to the proper refractive index $n^\mathrm{(A)}$. Thus, there is an extremely small difference between the total momentum density of the MP state of light in the L frame and the Minkowski momentum density $\mathbf{G}_\mathrm{M}^\mathrm{(L)}=\mathbf{D}^\mathrm{(L)}\times\mathbf{B}^\mathrm{(L)}$ that is proportional to $n^\mathrm{(L)}$. This result differs from the total momentum density of the coupled MP state of light in the L frame as presented in our previous works, where we effectively assumed that $n^\mathrm{(A)}$ is exactly equal to $n^\mathrm{(L)}$. The equality between the Minkowski and MP momentum densities applies \emph{exactly} in the A frame, where the local atomic velocity is zero. However, since the atomic velocity is extremely small in the L frame of realistic materials, the difference of $n^\mathrm{(L)}$ and $n^\mathrm{(A)}$ is also extremely small.

In the limit of large atomic mass density of the medium, when $\mathbf{v}_\mathrm{a}\rightarrow0$ and $n^\mathrm{(A)}\rightarrow n^\mathrm{(L)}$, the energy and momentum densities in Eqs.~\eqref{eq:mdwenergy}--\eqref{eq:mpmomentum} are in full agreement with the approximations that we have used in our previous works \cite{Partanen2017c,Partanen2019a}. The MP SEM tensor then obtains in the L frame exactly the same form as given in Eq.~(B4) in Appendix B of our original work in Ref.~\cite{Partanen2017c}. In the OCD simulations of our previous works \cite{Partanen2017c,Partanen2017e,Partanen2018a}, these approximations have been verified within the numerical accuracy of 7 digits. However, one must remember that both the numerical accuracy and the approximation $n^\mathrm{(A)}=n^\mathrm{(L)}$ are much more accurate than the approximation of a nondispersive medium for any realistic material.

\section{\label{sec:conclusions}Conclusions}

In conclusion, we have presented the Lagrangian formulation of the MP theory of light. Starting from the well-known Lagrangian densities of the field and the medium, the present work provides a solid field-theoretical foundation for the description of propagation of light in a medium. We have shown starting from the known Lagrangian densities how the optical force obtains \emph{ab initio} an unambiguous expression that couples the dynamics of the medium to the dynamics of the electromagnetic field. Thus, our work presents the derivation of the Abraham force from the first principles. Accordingly, the atomic MDW arises from the perturbation of the mass density of the medium under the influence of the optical force. The atomic mass density and the atomic velocity of the MDW satisfy the Newtonian equation of motion. This contrasts to the conventional theories of the propagation of light in a medium, where the effects of the optical force density on the dynamical state of the nondispersive medium are neglected. If the optical force and the resulting atomic MDW are neglected, the conservation laws of the total SEM tensor of the coupled state of the field and the medium become unavoidably violated. This is a strong argument for the MP theory of light to be the unique physically correct theory of light in a medium.

The Euler-Lagrange equations of the field and the medium include implicit symmetric bi-directional coupling of the subsystems that fulfills the law of action and counteraction. We have been able to demonstrate for a Gaussian light pulse the exact analytical solutions of the coupled dynamical equations for both the field and the medium. The mathematical methods developed to describe this bi-directionally coupled dynamical system may find applications far outside the present problem. We have also presented exact mathematical treatment of the kinetic energy of the MDW associated with light. The kinetic energy of the MDW is extremely small for light pulses having field intensities below the irradiation damage threshold of realistic materials. In contrast, the rest energy of atoms moving with the MDW gives rise to large energy flux in a transparent solid, such as silicon, as demonstrated in Fig.~\ref{fig:graphs}.

\begin{acknowledgments}
This work has been funded by the Academy of Finland under Contract No.~318197 and H2020 Marie Sk\l{}odowska-Curie Actions (MSCA) individual fellowship DynaLight under Contract No.~846218. Mathematica has been extensively used to verify the equations of the present work.
\end{acknowledgments}

\appendix

\section{\label{apx:materialparameters}Relation of material parameters in the A frame and in the L frame}

In this section, we describe the transformations of material parameters between the A frame and the L frame. First, we note that in the general inertial frame, G frame, the local velocity of light is presented in terms of the field quantities by the well-known relation as \cite{Landau1984}
\begin{equation}
 \mathbf{v}_\mathrm{l}=\frac{\mathbf{E}\times\mathbf{H}}{\frac{1}{2}(\mathbf{E}\cdot\mathbf{D}+\mathbf{H}\cdot\mathbf{B})}.
\label{eq:velocityoflight}
\end{equation}
The length of this velocity vector of light also defines the refractive index in the G frame as $|\mathbf{v}_\mathrm{l}|=c/n^\mathrm{(G)}$. From the perspective of atoms, there are two special inertial frames as described in Sec.~\ref{sec:constitutive}. The first is the A frame, which is a local inertial frame comoving with the atoms. The second is the L frame, where the atomic velocity is zero in the absence of the optical force. Below, we describe the transformations of material parameters between these two inertial frames.

The relation of the refractive index $n^\mathrm{(L)}$ of the L frame to the proper refractive index $n^\mathrm{(A)}$, defined in the A frame, is given by
\begin{align}
 &\mathbf{v}_\mathrm{l}^\mathrm{(A)}=\mathbf{v}_\mathrm{l}^\mathrm{(L)}\ominus\mathbf{v}_\mathrm{a}^\mathrm{(L)},\nonumber\\
&|\mathbf{v}_\mathrm{l}^\mathrm{(A)}|=c/n^\mathrm{(A)},
\hspace{0.4cm}|\mathbf{v}_\mathrm{l}^\mathrm{(L)}|=c/n^\mathrm{(L)},
\label{eq:ncondition}
\end{align}
where $\ominus$ denotes the relativistic velocity subtraction, defined for two arbitrary velocities $\mathbf{v}_1$ and $\mathbf{v}_2$ by the conventional relation \cite{Jackson1999}
\begin{equation}
 \mathbf{v}_1\ominus\mathbf{v}_2=\frac{1}{1-\frac{\mathbf{v}_1\cdot\mathbf{v}_2}{c^2}}\Big(\mathbf{v}_1-\frac{\mathbf{v}_\mathrm{2,\perp}}{\gamma_1}-\mathbf{v}_\mathrm{2,\parallel}\Big).
\label{eq:subtraction}
\end{equation}
Here $\parallel$ and $\perp$ denote the components parallel and perpendicular to $\mathbf{v}_1$, and $\gamma_1$ is the Lorentz factor corresponding to $\mathbf{v}_1$. Since, in the L frame, the atomic velocities are driven forward by the optical force that points in the local propagation direction of light, $\mathbf{v}_\mathrm{l}$ and $\mathbf{v}_\mathrm{a}$ are parallel, and Eqs.~\eqref{eq:ncondition} and \eqref{eq:subtraction} lead to the transformations of the refractive index, given by
\begin{equation}
n^\mathrm{(A)}=\frac{\displaystyle1-\frac{v_\mathrm{a}^\mathrm{(L)}}{n^\mathrm{(L)}c}}{\displaystyle\frac{1}{n^\mathrm{(L)}}-\frac{v_\mathrm{a}^\mathrm{(L)}}{c}},
\hspace{0.5cm}n^\mathrm{(L)}=\frac{\displaystyle1+\frac{v_\mathrm{a}^\mathrm{(L)}}{n^\mathrm{(A)}c}}{\displaystyle\frac{1}{n^\mathrm{(A)}}+\frac{v_\mathrm{a}^\mathrm{(L)}}{c}}.
\label{eq:refractiveindex}
\end{equation}

The relations between the permittivity and permeability in the A frame to the corresponding quantities in the L frame are obtained from the relation of the refractive indices in Eq.~\eqref{eq:refractiveindex} by using equations  $n^\mathrm{(L)}=c\sqrt{\mu^\mathrm{(L)}\varepsilon^\mathrm{(L)}}$ and $\sqrt{\mu^\mathrm{(L)}/\varepsilon^\mathrm{(L)}}=|\mathbf{E}^\mathrm{(L)}|/|\mathbf{H}^\mathrm{(L)}|=|\mathbf{E}^\mathrm{(A)}|/|\mathbf{H}^\mathrm{(A)}|=\sqrt{\mu^\mathrm{(A)}/\varepsilon^\mathrm{(A)}}$. The last equation, the constancy of the wave impedance, follows directly from the Lorentz transformation of the fields described in Appendix \ref{apx:Lorentz} when applied to the transformation from the A frame to the L frame. Therefore, the transformations of the permittivity and permeability of the medium are obtained together with the transformations of the refractive index in Eq.~\eqref{eq:refractiveindex} as
\begin{align}
\varepsilon^\mathrm{(A)}=\sqrt{\frac{\varepsilon^\mathrm{(L)}}{\mu^\mathrm{(L)}}}\frac{n^\mathrm{(A)}}{c},
\hspace{0.5cm}\varepsilon^\mathrm{(L)}=\sqrt{\frac{\varepsilon}{\mu}}\frac{n^\mathrm{(L)}}{c},\nonumber\\
\mu^\mathrm{(A)}=\sqrt{\frac{\mu^\mathrm{(L)}}{\varepsilon^\mathrm{(L)}}}\frac{n^\mathrm{(A)}}{c},
\hspace{0.5cm}\mu^\mathrm{(L)}=\sqrt{\frac{\mu}{\varepsilon}}\frac{n^\mathrm{(L)}}{c}.
\label{eq:permittivitypermeability}
\end{align}

Note that the relations in Eqs.~\eqref{eq:refractiveindex} and \eqref{eq:permittivitypermeability} can be generalized to describe the transformations of material parameters between any inertial frames whose relative velocity is parallel to the velocity of light. In this generalization, the atomic velocity is replaced by the relative velocity of the inertial frames. If this boost velocity is not parallel to the velocity of light, then the fields $\mathbf{D}$ and $\mathbf{E}$ are not generally parallel to each other. The same applies to the fields $\mathbf{B}$ and $\mathbf{H}$. Thus, the constitutive relations can generally be presented only as equations, where the scalar permittivity and permeability are replaced by matrices. The forms of these matrices could be derived from the Lorentz transformations of the fields, described in Appendix \ref{apx:Lorentz}.

\section{\label{apx:Lorentz}Lorentz transformations of the electromagnetic field and the atomic MDW quantities}

As presented in Ref,~\cite{Partanen2019a}, the total SEM tensor of the coupled MP state of the field and the MDW transforms in a Lorentz covariant way from the L frame to an arbitrary inertial frame moving with velocity $\mathbf{v}$ with respect to the L frame. In the MP theory, we utilize the Lorentz transformation of the electric and magnetic fields of the standard Minkowski form, given by \cite{Kemp2017,Penfield1967}
\begin{align}
 \mathbf{E}'&=\mathbf{E}_\parallel+\gamma(\mathbf{E}_\perp+\mathbf{v}\times\mathbf{B}),\nonumber\\
 \mathbf{H}'&=\mathbf{H}_\parallel+\gamma(\mathbf{H}_\perp-\mathbf{v}\times\mathbf{D}),\nonumber\\
 \mathbf{D}'&=\mathbf{D}_\parallel+\gamma(\mathbf{D}_\perp+\frac{1}{c^2}\mathbf{v}\times\mathbf{H}),\nonumber\\
 \mathbf{B}'&=\mathbf{B}_\parallel+\gamma(\mathbf{B}_\perp-\frac{1}{c^2}\mathbf{v}\times\mathbf{E}).
 \label{eq:LorentzFields}
\end{align}
These relations are built in the Lorentz covariance of the electromagnetic field tensor $F^{\alpha\beta}$, the electromagnetic displacement tensor $\mathcal{D}^{\alpha\beta}$, and the related Maxwell's equations.

In addition, the MDW mass density, the velocity of light, the velocity of atoms, and the velocity of the L frame transform between arbitrary inertial frames as \cite{Partanen2019a}
\begin{align}
 \rho_\mathrm{MDW}' &=\frac{c^2-\mathbf{v}_\mathrm{l}\cdot\mathbf{v}}{c^2-(\mathbf{v}\ominus\mathbf{v}_\mathrm{a})\cdot\mathbf{v}}\rho_\mathrm{MDW},\nonumber\\
 \mathbf{v}_\mathrm{l}' &=-(\mathbf{v}\ominus\mathbf{v}_\mathrm{l}),\nonumber\\
 \mathbf{v}_\mathrm{a}' &=-(\mathbf{v}\ominus\mathbf{v}_\mathrm{a}),\nonumber\\
 \mathbf{v}_\mathrm{rel}' &=-(\mathbf{v}\ominus\mathbf{v}_\mathrm{rel}).
 \label{eq:LorentzMDW}
\end{align}
Equations \eqref{eq:LorentzMDW} present the essential difference between the MP theory of light in a medium and the conventional Minkowski SEM theory. The MP formulation is also different from other known formulations of electrodynamics, none of which presents the atomic MDW as an integral part of the total coupled state of light in a medium.

\end{document}